\documentclass[preprint,showpacs,preprintnumbers,amsmath,amssymb,superscriptaddress,10pt,nofootinbib]{revtex4}

\usepackage{dcolumn}
\usepackage{bm}
\usepackage{amssymb}

\usepackage{graphicx}
\usepackage{epstopdf}
\usepackage{latexsym}
\usepackage{times}
\usepackage{multirow}
\usepackage{natbib}
\usepackage{amssymb,amsmath}
\usepackage{color}
\usepackage{colordvi}
\usepackage{mathrsfs}

\usepackage[center]{subfigure}

\def\MNRAS{Mon. Not. Roy. Astron. Soc.}
\newcommand{\be}{\begin{equation}}
\newcommand{\ee}{\end{equation}}
\newcommand{\bea}{\begin{eqnarray}}
\newcommand{\ena}{\end{eqnarray}}

\begin{document}

\title{Testing Brans-Dicke gravity using the Einstein telescope }


\author{Xing Zhang}
\author{Jiming Yu}
\author{Tan Liu}
\author{Wen Zhao}
\email[]{wzhao7@ustc.edu.cn}
\affiliation{CAS Key Laboratory for Research in Galaxies and Cosmology, Department of Astronomy, University of Science and Technology of China, Hefei 230026, China}
\affiliation{School of Astronomy and Space Science, University of Science and Technology of China, Hefei 230026, China}
\author{Anzhong Wang}
\affiliation{Institute for Advanced Physics and Mathematics, Zhejiang University of Technology,  Hangzhou, Zhejiang 310032, China}
\affiliation{GCAP-CASPER, Department of Physics, Baylor University, Waco, Texas 76798-7316, USA}

\date{\today}


\begin{abstract}
{
Gravitational radiation is an excellent field for testing theories of gravity in strong gravitational fields. The current observations on the gravitational-wave (GW) bursts by LIGO have already placed  various constraints on the alternative theories of gravity. In this paper, we investigate the possible bounds which could be placed on the Brans-Dicke gravity using GW detection from inspiralling compact binaries with the proposed Einstein Telescope, a third-generation GW detector. We first calculate in details the waveforms of gravitational radiation in the lowest post-Newtonian approximation, including the tensor and scalar fields, which can be divided into the three polarization modes, i.e. ``plus mode", ``cross mode" and ``breathing mode". Applying the stationary phase approximation, we obtain their Fourier transforms, and derive the correction terms in amplitude, phase and polarization of GWs, relative to the corresponding results in General Relativity. Imposing the noise level of Einstein Telescope, we find that the GW detection from inspiralling compact binaries, composed of a neutron star and a black hole, can place stringent constraints on the Brans-Dicke gravity. The bound on the coupling constant $\omega_{\rm BD}$ depends on the mass, sky-position, inclination angle, polarization angle, luminosity distance, redshift distribution and total observed number $N_{\rm GW}$ of the binary systems. Taking into account all the burst events up to redshift $z=5$, we find that the bound could be $\omega_{\rm BD}\gtrsim 10^{6}\times(N_{\rm GW}/10^4)^{1/2}$. Even for the conservative estimation with $10^{4}$ observed events, the bound is still more than one order tighter than the current limit from Solar System experiments. So, we conclude that Einstein Telescope will provide a powerful platform to test alternative theories of gravity.
}

\end{abstract}

\pacs{98.70.Vc, 98.80.Cq, 04.30.-w}

\maketitle


\section{Introduction \label{section1}}

Since Einstein's General Relativity (GR) was proposed more than 100 years ago, a large number of experimental tests have been performed on various scales, from submillimeter scales tests in the laboratory, to the tests in Solar System and cosmological scales  \cite{will-book,will-review,Clifton2012p1,Stairs2003p5,Wex2014}. Even so, most of these efforts have focused on the gravitational effects in weak fields. Different from them, gravitational radiation provides an excellent opportunity to experimentally test  gravitational theories in the strong field regime. Since the observed gravitational waves (GWs) are always produced in either strong gravitational fields, extremely high energy scales, or the very early Universe, and are nearly freely propagating in the spacetime once generated, they encodes the clean information of these extreme conditions. Thus, a huge attention has been devoted to the detection of GWs. On September 14, 2015, the first direct GW signal,  GW150914,  was observed by LIGO, which marks the beginning of the era of GW astronomy \cite{gw150914}. Since then, various investigations on testing GR, including those from LIGO collaborations, have been carried out by utilizing the observed GW data \cite{gw150914,yunes2016,arzaon2016,maselli}.

Karl Popper argued that scientists can never truly ``prove" that a theory, including GR, is correct, but rather all we do is to disprove, or more accurately to constrain a hypothesis. The theory that remains and cannot be disproved by observations becomes the {\emph{status quo}} \cite{karl}. According to this argument, in order to test GR we much compare its predictions with alternative theories of gravity. So, the theoretical studies on gravitational radiations in various theories are highly desirable. For instance, in the previous work \cite{ppe}, the authors developed the parameterized post-Einsteinian (ppE) framework to describe the modifications of GWs in a wide class of gravitational theories.

In this paper, we will focus on Brans-Dicke (BD) gravity. As the simplest scalar-tensor gravity, BD gravity has been well studied and constrained in \textcolor{black}{ various tests} (see for instance \cite{bd-book1,bd-book2}). For the gravitational radiation of inspiralling compact binaries in BD gravity, Will et al. have calculated the gravitational waveforms by including the lowest order effects \cite{will-book,will1989,will1994}\footnote{These calculations have been extended to the higher post-Newtonian (PN) orders in the recent works \cite{lang2013,lang2014,bd-higher,Sennett:2016klh}.}. Similar calculations have also applied to some extended versions of BD gravity \cite{damour,will-massive,barx2014,cao,zhang}. However, for the gravitational waveforms, in these works the authors have only considered the phase correction terms in the  ``plus-mode" and ``cross-mode" of GWs. As well-known, for the compact systems, the predictions of gravitational radiation in BD gravity are different from those in GR in several aspects \cite{will-book}: \textcolor{black}{First, modifications} of the effective masses of the bodies, parameterized by the sensitivities $s_i$, alter the motion of two-body orbits, which induces the modification on the time dependence of the orbital frequency and GW frequency of the system. Second, in addition to the quadrupole gravitational radiation, in BD gravity the scalar field also emits  scalar radiations, including the monopole, dipole and quadrupole components. These radiations also modify the orbital evolution of the system and thence the GW frequency and amplitude. In this paper, we extend the previous calculations on the gravitational radiation of compact binary system in BD gravity, and derive the full waveforms of GWs by including the ``plus mode", ``cross mode" as well as the ``breathing mode". Employing the stationary phase approximation, we obtain the Fourier transforms of these components, and find that the contribution of scalar monopole radiation is negligible, and \textcolor{black}{the dipole and quadrupole scalar radiations} are suppressed by the BD parameter $\omega_{\rm BD}$ and/or the difference in sensitivities of two objects. The tensor quadrupole radiations are modified in both GW phases and amplitudes, which are significant in the low frequency range.

It is well-known that BD gravity reduces to GR in the limit $\omega_{\rm BD}\rightarrow\infty$. Many effects have been \textcolor{black}{devoted  to constrain the parameter} $\omega_{\rm BD}$ in various systems \cite{will-book,bd-book1,bd-book2}. Until now, the most stringent constraint is $\omega_{\rm BD}>4\times10^{4}$, which comes from the Cassini-Huygens experiment \cite{bound}. In the previous work \cite{will1994}, the authors showed that observations of inspiral binary systems from ground-based detectors of the type of the advanced LIGO could place a bound of $\omega_{\rm BD}\gtrsim2000$. If considering the LISA space interferometer, for a neutron star inspiralling into a $10^3$ $M_{\odot}$ black hole in the Virgo Cluster, a possible bound of $\omega_{\rm BD}\gtrsim3\times10^5$ could be placed in a two-year integration \cite{will-lisaa,will-lisa}. \textcolor{black}{Similar results are also derived in the previous works \cite{lisa2,lisa3}. In addition, if considering the observations of potential space-based DICEGO/BBO projects, the bound $\omega_{\rm BD}\gtrsim4\times 10^{8}$ could be placed in the far future \cite{bbo-decigo}. } In this paper, we shall apply similar analyses to the potential observations of Einstein Telescope (ET). Currently, ET is undergoing a design study as a third-generation ground-based GW observatory \cite{et}, which would be able to observe binary neutron \textcolor{black}{star systems} up to redshift $z\sim 2$ and the neutron-star/black-hole events up to $z\sim 8$. Comparing with
\textcolor{black}{the generation of the advanced LIGO, which is often referred to as the second generation},  ET has the following advantages: The noise power spectral density (PSD) of ET will be more than two orders smaller, while the lower cutoff frequency of ET will extend to $1$ Hz. Both factors will greatly improve the total number of inspiralling compact binaries, as well as the signal-to-noise ratio for the given target. So, we anticipate that BD gravity can be  well constrained by the potential observations of ET. \textcolor{black}{In the previous works \cite{et2,et3}, the authors found that, if considering ET, one GW event could place a bound of $\omega_{\rm BD}\gtrsim(10^{4}\sim10^{5}$). In this paper, we shall extend these analyses by combining multiple events, and considering all the modifications of GW waveforms.}

The outline of this paper is as follows. In Sec. \ref{section2} we calculate the gravitational waveforms of compact binary \textcolor{black}{systems in BD gravity, derive their Fourier transforms} by applying the stationary phase approximation, and then extend them \textcolor{black}{ to include} high PN terms. In Sec. \ref{section3}, we discuss the capabilities of \textcolor{black}{ET on} constraining BD gravity by taking into account a large number of GW events in a wide redshift range. In Sec. \ref{section4} we conclude
\textcolor{black}{the paper} with a summary of our main results.

Throughout this paper, the \textcolor{black}{signatures of metric are chosen as} $(-,+,+,+)$, and the Greek indices ($\mu,\nu,\cdots$) run over $0,1,2,3$. We choose the units in which $G=c=1$, where $G$ is the Newtonian gravitational constant, and $c$ is the speed of light in vacuum.


\section{Gravitational radiations in  scalar-tensor gravity \label{section2}}

\subsection{  \textcolor{black}{BD} gravity \label{section20}}

In the Jordan frame, the action of the general scalar-tensor gravity is given by \cite{will-book}
\bea\label{action}
I=\frac{1}{16\pi}\int \left[\phi R-\frac{\omega{(\phi)}}{\phi}g^{\mu\nu}\phi_{,\mu}\phi_{,\nu}+2\phi\lambda(\phi)\right]\sqrt{-g}d^4 x+I_{m}(g_{\mu\nu},q_{A}),
\ena
where $g_{\mu\nu}$ is the spacetime metric, $g$ is its determinant, $R$ is the Ricci scalar derived from this metric, $\phi$ is the scalar field, and $\omega(\phi)$ is the scalar-tensor coupling \textcolor{black}{function}, $\lambda(\phi)$ is the cosmological function. $I_{m}$ represents the matter action, which depends only on the matter fields $q_A$ and the metric $g_{\mu\nu}$, i.e. there is no direct interaction with the scalar field. In this paper, we restrict our attention to the massless \textcolor{black}{BD} theory, in which $\omega(\phi)=\omega_{\rm BD}$ is a constant, and $\lambda(\phi)=0$.

The field equations derived from the action of \textcolor{black}{BD} gravity are given by
\bea
R_{\mu\nu}-\frac{1}{2}g_{\mu\nu}R=\frac{8\pi}{\phi}T_{\mu\nu}+\frac{\omega_{\rm BD}}{\phi^2}\left(\phi_{,\mu}\phi_{,\nu}-\frac{1}{2}g_{\mu\nu}\phi_{,\rho}\phi^{,\rho}\right)+\frac{1}{\phi}(\phi_{;\mu\nu}-g_{\mu\nu}\Box_{g}\phi),
\ena
\bea
\Box_{g}\phi=\frac{1}{3+2\omega_{\rm BD}}\left(8\pi T-16\pi\phi\frac{\partial T}{\partial\phi}\right),
\ena
where $T_{\mu\nu}$ is the stress-energy tensor of matter and nongravitational fields, and $T\equiv g^{\alpha\beta}T_{\alpha\beta}$ is its trace. Throughout this paper, we use commas to denote ordinary derivatives, semicolons \textcolor{black}{ to} denote covariant derivatives, and \textcolor{black}{$\Box_{g}\equiv g^{\alpha\beta}\nabla_{\alpha}\nabla_{\beta}$} \textcolor{black}{to represent} the d'Alembertian with indices raised by the metric \textcolor{black}{$g^{\mu\nu}$}. Here, we should mention that in the general Jordan frame, the quantity $\partial T/\partial \phi$ is not present. But for the gravitationally bound bodies, as to be shown below, it will be present in the field \textcolor{black}{equations.}

In order to discuss the gravitational radiation, we assume that far away from the sources, the metric $g_{\mu\nu}$ reduces to the Minkowski metric $\eta_{\mu\nu}$, and the scalar field $\phi$ tends to \textcolor{black}{its}
 cosmological value $\phi_0$. Thus, we can define the perturbations in the
far-zone as follows, \bea
h_{\mu\nu}=g_{\mu\nu}-\eta_{\mu\nu},~~~\varphi=\phi-\phi_0, \ena
\bea
\theta^{\mu\nu}=h^{\mu\nu}-\frac{1}{2}h\eta^{\mu\nu}-(\varphi/\phi_0)\eta^{\mu\nu},
\ena where $\varphi$ is the
perturbation of \textcolor{black}{the} scalar field $\phi$ about its asymptotic
cosmological value $\phi_0$. Note that in another version of \textcolor{black}{the} field equations \cite{will-book2,lang2013}, an auxiliary metric $\tilde{g}_{\mu\nu}$ is introduced, which relates to the physical metric $g_{\mu\nu}$ by the conformal transformation $\tilde{g}_{\mu\nu}\equiv(\phi/\phi_0)g_{\mu\nu}$, and a ``gothic" version of this \textcolor{black}{metric}, $\tilde{\mathfrak{g}}^{\mu \nu} \equiv \sqrt{-\tilde{g}}\tilde{g}^{\mu \nu}$. In the weak field \textcolor{black}{approximations}, it can be proved that $\theta^{\mu\nu}=\eta^{\mu\nu}-\tilde{\mathfrak{g}}^{\mu \nu}$. Following the previous works \cite{will-book,will1989,will1994}, in this paper we shall use the \textcolor{black}{ quantities} $\theta^{\mu\nu}$ and $\varphi$. Choosing \textcolor{black}{the} harmonic gauge in which
$\theta^{\mu\nu}_{,~\nu}=0$, we can rewrite the field equations for
BD theory in the form \bea \label{wave-equation}
\Box_{\eta}\theta^{\mu\nu}=-16\pi\tau^{\mu\nu},~~~\Box_{\eta}\varphi=-8\pi
\tau_s, \ena where the sources terms $\tau^{\mu\nu}$ and $\tau_s$ are explicitly given in \cite{will1989,will-book2,lang2013}, and $\tau^{\mu\nu}$ satisfies the conservation laws $\tau^{\mu\nu}_{,~\nu}=0$ because of the Bianchi \textcolor{black}{identity}. Note that the indices \textcolor{black}{of} $\theta^{\mu\nu}$ and $\varphi_{,\mu}$ \textcolor{black}{will be lowered and raised by ${\bf \eta}_{\mu\nu}$ and ${\bf \eta}^{\mu\nu}$}.

\subsection{ \textcolor{black}{Evolution of binary systems in BD} gravity \label{section2a}}

Now, let us \textcolor{black}{turn to consider } a realistic source, which \textcolor{black}{is made } of two compact objects. Since the compact \textcolor{black}{system} is gravitationally bound, its total mass depends on its internal gravitational energy, which in turn depends on the effective local value of  \textcolor{black}{the} scalar field $\phi$ in the vicinity of the body. Eardley found that these effects could be accounted for by  \textcolor{black}{simply replacing} the constant inertial mass of the object in the distributional stress-energy tensor of the ``crude" approach  \textcolor{black}{by} a function of the scalar field $\phi$, namely $m_i(\phi)$ ($i=1, 2$) \cite{eardley}. Thus, the matter action in Eq. (\ref{action}) becomes
\bea
I_{m}=-\sum_{i=1,2}\int m_i(\phi) d\tau_i,
\ena
where $\tau_i$  \textcolor{black}{denotes the } proper time along the trajectory of  \textcolor{black}{the} object $i$. These modifications depend on the internal structure of the bodies and \textcolor{black}{ the theory of gravity}. We expand $m_i(\phi)$ about the asymptotic value $\phi_0$ as follows,
\bea
m_i(\phi)=m_i\left[1+s_i\left(\frac{\phi}{\phi_0}\right)+\frac{1}{2}(s_i^2+s'_i-s_i)\left(\frac{\phi}{\phi_0}\right)^2+O\left(\frac{\phi}{\phi_0}\right)^3\right],
\ena
where $m_i\equiv m_i(\phi_0)$,  \textcolor{black}{and the sensitivity $s_i$ and its derivative $s'_i$ } are defined as
\bea
s_i\equiv \left(\frac{d \ln m_i(\phi)}{d \ln \phi}\right)_{\phi=\phi_0},~~s'_i\equiv \left(\frac{d^2 \ln m_i(\phi)}{d (\ln \phi)^2}\right)_{\phi=\phi_0}.
\ena
The sensitivities $s_i$ roughly measure the gravitational binding energy per unit mass. This effect violates the Strong Equivalence Principle,  in \textcolor{black}{ the sense } that the motion of such bodies now depends
\textcolor{black}{ on} their internal structure (apart from tidal interactions). In \textcolor{black}{BD} gravity, for \textcolor{black}{ white dwarfs we have  $s \simeq 0$}, for neutron stars $s\approx0.1-0.2$ \cite{will1989}, and for black holes $s=0.5$ \cite{hawking1972}.

The stress-energy tensor in this system is given by
\bea
T^{\mu\nu}=(-g)^{-1/2}\sum_{i=1,2}m_i(\phi)u_i^{\mu}u_i^{\nu}(u_i^{0})^{-1}\delta^3({\bf x}-{\bf x}_i),
\ena
\bea
\frac{\partial T}{\partial \phi}=-(-g)^{-1/2}\sum_{i=1,2}\frac{\partial m_i(\phi)}{\partial \phi}(u_i^{0})^{-1}\delta^3({\bf x}-{\bf x}_i),
\ena
where $u_i^{\mu}$ is the four-velocity of \textcolor{black}{the object}  $i$.

In this system, we treat the objects as point-like, with masses $m_1$ and $m_2$, and positions ${\bf x_1}$ and ${\bf x_2}$, \textcolor{black}{respectively}. From the post-Newtonian
equations of motion \cite{will-book}, in the center-of-mass frame, it was shown \textcolor{black}{ that} the dynamics in the Newtonian limit reduces to a one-body system with \textcolor{black}{a mass equal to the}
reduced mass $\mu=m_1m_2/(m_1+m_2)$, and \textcolor{black}{the} equation of motion \cite{will1989}
\bea\label{dxdt}
d^2{\bf x}/dt^2=-\mathcal{G} m{\bf x}/r^3,
\ena
where $m=m_1+m_2$ is the total mass and ${\bf x}={\bf x_2}-{\bf x_1}$ the relative coordinate. The parameter ${\mathcal G}$ is defined \textcolor{black}{as,}
\bea
\mathcal{G}=1-\xi(s_1+s_2-2s_1s_2),~~~\xi=(2+\omega_{\rm BD})^{-1}.
\ena

In this paper, we consider only the case of quasi-circular orbits (that is, circular, apart from an adiabatic inspiral). Then the orbital frequency $\omega_s$ is related to the orbital radius $r$ by $v^2=\mathcal{G}m/r$ with the orbital velocity $v=\omega_s r$. So, we have Kepler's third law
\bea\label{omega-s}
\omega_s=(\mathcal{G}m/r^3)^{1/2},
\ena
and the orbit period $P_s$ is given by $P_s=2\pi/\omega_s$. The energy of the system is given by
\bea
E=-(1/2)\mathcal{G}\mu m/r.
\ena

For \textcolor{black}{a} compact binary system, the dissipation of \textcolor{black}{its} total energy is caused by the emission of gravitational \textcolor{black}{ radiations}. In \textcolor{black}{BD} gravity, the rate of energy loss for a quasi-circular two-body orbit is given by \cite{will1989,will-book}
\bea\label{dEdt}
\frac{d E}{dt}=-\left\langle\frac{8}{15}\frac{\mu^2m^2}{r^4}(12\kappa v^2+\frac{5}{8}\kappa_D\mathcal{S}^2)\right\rangle,
\ena
where the angular brackets denote an orbital average, and the coefficients are given by
\bea
\kappa=\mathcal{G}^2(1-\frac{1}{2}\xi+\frac{1}{12}\xi\Gamma^2),~~
\kappa_D=2\mathcal{G}^2\xi,~~
\mathcal{S}=s_1-s_2,~~
\Gamma=1-2(m_1s_2+m_2s_1)/m.
\ena
The first term in Eq. (\ref{dEdt}) \textcolor{black}{represents} the combined effects \textcolor{black}{of the quadrupole and monopole radiations}, and the second term is the contribution of the dipole radiation. If $\kappa\rightarrow1$ and $\kappa_{D}\rightarrow0$, it \textcolor{black}{reduces to that } of GR.

\subsection{Gravitational waveforms in \textcolor{black}{BD} gravity \label{section2b}}

The gravitational radiations can be derived by solving the wave equations \textcolor{black}{of}  Eq. (\ref{wave-equation}).
For a binary orbit, to leading order \textcolor{black}{of}  $v^2\sim m/r$, the solutions of \textcolor{black}{the} spatial components of the perturbations are given by \cite{will1989,will-book}, \bea
\theta^{ij}&=&2(1-\xi/2){d_{\rm L}}^{-1}(d^2/dt^2)\sum_{k=1,2}m_k(\phi)
x_{k}^{i}x_{k}^j
=(4\mu/d_{\rm L})(1-\xi/2)(v^iv^j-\mathcal{G}mx^ix^j/r^3),\\
\varphi/\phi_0&=&\xi(\mu/d_{\rm L})\left\{\Gamma[({\bf {\hat{N}}\cdot
v})^2-\mathcal{G}m({\bf {\hat{N}}\cdot
x})^2/r^3]-(\mathcal{G}\Gamma+2\Lambda)m/r-2\mathcal{S}({\bf
{\hat{N}}\cdot v}) \right\},\ena where the parameter $\Lambda$ is given by $\Lambda=1-s_1-s_2$, $d_{\rm L}$ is the luminosity distance of the
observer, and ${\bf {\hat{N}}}$ is \textcolor{black}{the} direction unit vector of $d_{\rm L}$. In the flat Friedmann-Lema\^{\i}tre-Robertson-Walker universe, the luminosity distance is calculate by \cite{weinberg}
\bea
d_{\rm L}(z)=(1+z)\int_0^z\frac{dz'}{H(z')},
\ena
where $z$ is the redshift, and $H(z)$ is the Hubble parameter. In the spatial flat $\Lambda$CDM model, it is given by \bea H(z)=H_0[\Omega_m(1+z)^3+\Omega_{\Lambda}]^{1/2}.\ena
Throughout this paper, we adopt a fiducial cosmological model with \textcolor{black}{the}  following values \textcolor{black}{of} the parameters \cite{planck}:
\bea
\Omega_m=0.314,~~~\Omega_{\Lambda}=0.686,~~~\Omega_k=0,~~~H_0=67.3{\rm km~s^{-1}Mpc^{-1}}.
\ena

The perturbation of the metric is obtained by utilizing the relation \footnote{There is a typo in the formulae (2.7) of Ref. \cite{will1994}, where $\bar{h}^{ij}$ should be replaced by $h^{ij}$.}
\bea\label{hij}
h^{\mu\nu}=\theta^{\mu\nu}-\eta^{\mu\nu}\theta/2-(\varphi/\phi_0)\eta^{\mu\nu}. \ena
A gravitational-wave detector \textcolor{black}{measures} the separation $\xi^{i}$ between \textcolor{black}{ the two} test masses. If the distance between \textcolor{black}{them} is small compared to the wavelength of GWs, and the \textcolor{black} {test masses move} slowly, the separation obeys the equation
${d^2}\xi^i/{dt^2}=-R^{0i0j}\xi^{j}$ \cite{MTW}. The components of the Riemann tensor $R^{0i0j}$ measured by a
detector can be shown to be given by \cite{gw-book,MTW} \bea
R^{0i0j}=\frac{1}{2}(\partial^i\partial^0h^{0j}+\partial^j\partial^0h^{0i}-\partial^0\partial^0h^{ij}-\partial^i\partial^jh^{00})
\equiv-\frac{1}{2}\frac{d^2}{dt^2}{\textrm{h}^{ij}},\ena where we have defined the effective gravitational waveform $\textrm{h}^{ij}$.
Using Eq. (\ref{hij}) and the relation $\partial^i(\varphi/\phi_0)=-\partial_0(\varphi/\phi_0)\hat{N}^i$,
we derive that \bea \label{text-hij}\textrm{h}^{ij}=
\theta_{\rm TT}^{ij}-(\varphi/\phi_0)(\delta^{ij}-\hat{N}^{i}\hat{N}^{j}),\ena
where $\text {TT}$ denotes the transverse-traceless projection. Note that
the full gravitational waveform is transverse but not traceless
because of the presence of the scalar contribution in Eq. (\ref{text-hij}). For quasi-circular orbits, by employing the relation $v^2=\mathcal{G}m/r$, the waveform \textcolor{black}{of}  Eq. (\ref{text-hij}) becomes \cite{will1994,lang2013} \footnote{There is a typo in the formulae (2.9) and (2.10c) of Ref. \cite{will1994}, and the similar typo also appears in Eq. (41) of Ref. \cite{yunes2012}.}
\bea\label{rm-h-ij}
\textrm{h}^{ij}=\frac{2\mu}{d_{\rm L}}\left[Q_{TT}^{ij}+S(\delta^{ij}-\hat{N}^i\hat{N}^j)\right],\ena
\bea
Q^{ij}=2(1-\frac{1}{2}\xi)\frac{\mathcal{G}m}{r}(\hat{\lambda}^i\hat{\lambda}^j-\hat{n}^i\hat{n}^j),\ena
\bea S=-\frac{1}{2}\xi\left\{\frac{\Gamma\mathcal{G}m}{r}[({\bf
{\hat{N}}}\cdot{\bf{\hat{\lambda}}})^2-({\bf
{\hat{N}}}\cdot{\bf{\hat{n}}})^2]-(\mathcal{G}\Gamma+2\Lambda)\frac{m}{r}-2\mathcal{S}(\frac{\mathcal{G}m}{r})^{1/2}{\bf
{\hat{N}}}\cdot{\bf{\hat{\lambda}}}\right\},\ena
where we have defined the unit vectors ${\bf{\hat{n}}}\equiv {\bf x}/r$ and ${\bf{\hat{\lambda}}}\equiv{\bf v}/v$.

In metric theories of gravity, up to six degrees of freedom are allowed \cite{will-book}. \textcolor{black}{In addtion to}  the ``plus mode" (denoted as $+$, the real part of \textcolor{black}{the Weyl tensor component } $\Psi_4$) and ``cross mode" (denoted as $\times$, the imaginary part of \textcolor{black}{the Weyl tensor component $\Psi_4$)}, they include a scalar ``breathing mode" (denoted as $b$, the traceless \textcolor{black}{part of the} Ricci tensor, $\Phi_{22}$), a scalar longitudinal mode (denoted as $L$, the Weyl tensor \textcolor{black}{component} $\Psi_2$), and two vectorial modes (denoted as $x$ and $y$, the real and imaginary parts of \textcolor{black}{the Weyl tensor component } $\Psi_{3}$, respectively ). So, in general, the full (effective) \textcolor{black}{metric perturbations take} the form
\bea
{\rm h}_{ij}=h_{+} {\rm e}^{+}_{ij}+h_{\times} {\rm e}^{\times}_{ij}+h_{b} {\rm e}^{b}_{ij}+h_{L} {\rm e}^{L}_{ij}+h_{x} {\rm e}^{x}_{ij}+h_{y} {\rm e}^{y}_{ij},
\ena
where the polarization tensors are defined as
\bea
{\rm e}^{+}_{ij}&=&\hat{e}_x\otimes\hat{e}_x-\hat{e}_y\otimes\hat{e}_y,~~{\rm e}^{\times}_{ij}=\hat{e}_x\otimes\hat{e}_y+\hat{e}_y\otimes\hat{e}_x, \\
{\rm e}^{b}_{ij}&=&\hat{e}_x\otimes\hat{e}_x+\hat{e}_y\otimes\hat{e}_y,~~{\rm e}^{L}_{ij}=\hat{e}_z\otimes\hat{e}_z,\\
{\rm e}^{x}_{ij}&=&\hat{e}_x\otimes\hat{e}_z+\hat{e}_z\otimes\hat{e}_x,~~{\rm e}^{y}_{ij}=\hat{e}_y\otimes\hat{e}_z+\hat{e}_z\otimes\hat{e}_y.
\ena

In the ${\rm E}(2)$ classification for \textcolor{black}{GWs}, the massless scalar-tensor theories (including BD gravity) are of Class ${\rm N}_3$, i.e. the nonzero components are $h_{+}$, $h_{\times}$ and $h_b$ \cite{will-book}. In Appendix A, we proved that the first term of $\textrm{h}^{ij}$ in Eq. (\ref{rm-h-ij}) corresponds to the ``plus" and
``cross" polarization modes of the \textcolor{black}{GWs}. From Eq. (\ref{rm-h-ij}), we observe that these two terms are given by
\bea h_{+}(t)&=&-\frac{4\mathcal{G}\mu
m}{d_{\rm L} r}(1-\frac{1}{2}\xi)\frac{1+\cos^2\iota}{2}\cos2\Phi(t),\\
h_{\times}(t)&=&-\frac{4\mathcal{G}\mu
m}{d_{\rm L} r}(1-\frac{1}{2}\xi)\cos\iota\sin2\Phi(t),\ena
where $\iota$ is the  \textcolor{black}{inclination angle of  the binary orbital angular momentum along the line of sight}. \textcolor{black}{The polarization angle is calculated by $\Phi(t)=\int_{t_0}^{t} \omega_s(t') dt'+\Phi_0$, where $\Phi_0$ is the initial phase at $t=t_0$.} In BD gravity, from the relation  \textcolor{black}{of} (\ref{omega-s}), we have
\bea
\frac{\mu m}{r}=M_c^{5/3}(2\pi
f_s)^{2/3}\mathcal{G}^{-1/3},
\ena
where $M_c=\mu^{3/5}m^{2/5}$ is the chirp mass. Thus, these two components can be rewritten as
\bea
h_{+}(t)&=& -\frac{4\beta}{d_{\rm L}}M_c^{5/3}(2\pi
f_s)^{2/3}\frac{1+\cos^2\iota}{2}\cos2\Phi(t),\label{hplus}\\
h_{\times}(t)&=&-\frac{4\beta}{d_{\rm L}}M_c^{5/3}(2\pi
f_s)^{2/3}{\cos\iota}\sin2\Phi(t), \label{h_cross}
\ena where $\beta\equiv(1-\frac{1}{2}\xi)\mathcal{G}^{2/3}$ is the correction factor in BD gravity. In the case \textcolor{black}{ $\beta=1$, these results reduce to those of GR } \cite{gw-book}.

The second term in Eq. (\ref{rm-h-ij}) exactly corresponds to the breathing mode of  \textcolor{black}{GWs, which can be written as the sum of three terms},
 \bea
h_{b}(t)=\frac{2\mu}{d_{\rm L}}S\equiv h_{b1}+h_{b2}+h_{b3}, \label{hb}\ena
where
\bea h_{b1}(t)=-\frac{\mu m}{d_{\rm L} r}({\xi
\Gamma\mathcal{G}})\sin^2\iota\cos2\Phi(t), ~~~
h_{b2}(t)=\frac{\mu m}{d_{\rm L}r}(\mathcal{G}\Gamma+2\Lambda),~~~
h_{b3}(t)=\frac{2\mu}{d_{\rm L}}(\xi\mathcal{S})({\mathcal{G}
m}/{r})^{1/2}\sin\iota\cos\Phi(t).
\ena
Using the relation  \textcolor{black}{ (\ref{omega-s})}, they can be rewritten as
\bea
h_{b1}(t)&=&-\frac{{\xi\Gamma\mathcal{G}}}{d_{\rm L}}M_c^{5/3}(2\pi
f_s)^{2/3}\mathcal{G}^{-1/3}\sin^2\iota\cos2\Phi(t), \label{hb1}\\
h_{b2}(t)&=&\frac{\mathcal{G}\Gamma+2\Lambda}{d_{\rm L}}M_c^{5/3}(2\pi
f_s)^{2/3}\mathcal{G}^{-1/3},\label{hb2}\\
h_{b3}(t)&=&\frac{2\xi\mathcal{S}}{d_{\rm L}}M_c^{5/3}(2\pi
f_s/m)^{1/3}\sin\iota\cos\Phi(t).\label{hb3}
\ena

\subsection{ Waveforms in the stationary phase approximation\label{section2c}}

To compute the Fisher information matrix we would need the Fourier transform $\tilde{h}(f)$ of the signal $h(t)$. During the inspiral, the change in orbital frequency over a single period is negligible, and it is possible to apply a stationary phase approximation (SPA) to compute the Fourier transformation. Given a function $B(t)=2A(t)\cos\phi(t)$, where $d\ln A/dt\ll d\phi(t)/dt$ and $|d^2\phi/dt^2|\ll(d\phi/dt)^2$, the SPA provides the following estimate of the Fourier transform $\tilde{B}(f)$ (see, for instance, \cite{gw-book}):
\bea\label{general-ft}
\tilde{B}(f)\simeq \frac{A(t_f)}{\sqrt{\dot{F}(t_f)}} e^{i[\Psi_f(t_f)-\pi/4]},~~f\ge0,
\ena
where $\Psi_f(t)\equiv 2\pi ft-\phi(t)$, $2\pi F(t)\equiv d\phi/dt$. In this equation $t_f$ is defined as the time at which $F(t_f)=f$ and $\Psi_f(t_f)$ is the value of $\Psi_f(t)$ at $t=t_f$. We first calculate the evolution of the frequency $f_s\equiv w_s/2\pi$ in BD gravity.
From the evolution equations  \textcolor{black}{ (\ref{dxdt}), (\ref{omega-s}) and (\ref{dEdt})}, we derive that
\bea
\dot{f_s}=\frac{48\mu{\mathcal{G}}^{1/2}}{5\pi m^3}\left(\frac{m}{r}\right)^{11/2}\left(\kappa+\frac{5}{96}\frac{\kappa_D}{\mathcal{G}}\frac{r}{m}\mathcal{S}^2\right),
\ena
 \textcolor{black}{ from which we find} that
\bea
(2\pi\mathcal{M}_c f_s)^{-8/3}[1-(4/5)b\eta^{2/5}(2\pi\mathcal{M}_c f_s)^{-2/3}]=(256/5)(t_c-t)/\mathcal{M}_c,
\ena
where $\eta\equiv\mu/m$ is the symmetric mass ratio, $t_c$ is the time at which $f_s\rightarrow\infty$. We have defined the quantities,
\bea
\mathcal{M}_c\equiv(\kappa^{3/5}/\mathcal{G}^{4/5})\eta^{3/5}m,~~~b\equiv(5/96)(\kappa^{-3/5}\mathcal{G}^{-6/5})\kappa_D\mathcal{S}^2.
\ena
In the case  \textcolor{black}{ $\kappa\rightarrow1$ and $\mathcal{G}\rightarrow1$}, we find that $\mathcal{M}_c$ reduces to the chirp mass $M_c$. In BD gravity with $\xi\ll1$ (i.e. $\omega_{\rm BD}\gg 1$) and $\mathcal{S}\lesssim 1$, we always have $b\ll1$. Taking into account the fact that \cite{will1994}
\bea
z\equiv b\eta^{2/5}(2\pi\mathcal{M}_c f_s)^{-2/3} \le 5\times 10^{-3}\left(\frac{500}{\omega_{\rm BD}}\right)\left(\frac{\mathcal{S}}{0.5}\right)^2\left(\frac{M_\odot}{\mathcal{M}_c}\right)\left(\frac{30{\rm Hz}}{f_s}\right)^{2/3},
\ena
up to the first order of $z$, we obtain the relation between $f_s$ and $t$,
\bea\label{omega-s-t}
\omega_s=2\pi f_s=\frac{1}{\mathcal{M}_c}\left(\frac{256(t_c-t)}{5\mathcal{M}_c}\right)^{-3/8}\left[1-\frac{3}{10}b\eta^{2/5}
\left(\frac{256(t_c-t)}{5\mathcal{M}_c}\right)^{1/4}\right].
\ena

Now, let us focus on the Fourier transformation of the ``plus mode" by utilizing the result  \textcolor{black}{ (\ref{hplus})}. Since the amplitude varies slowly in comparison with the phase $2\Phi(t)$, the stationary point $t_*(f)$ is determined by the condition $2\pi f=2\dot\Phi(t_*)=4\pi f_s(t_*)$, i.e. $f=2f_s(t_*)$, which expresses the  \textcolor{black}{ fact} that the largest contribution to the Fourier component $\tilde{h}_+(f)$ with a given $f$ is obtained for the value of $t$ such that the chirping frequency $f$ is equal to $2f_s$. Using the relations  \textcolor{black}{given}  in (\ref{general-ft}) and (\ref{omega-s-t}),  \textcolor{black}{ and} from (\ref{hplus}) we derive that
\bea
\tilde{h}_+(f)&=&\sqrt{\frac{5}{24}}\frac{1}{\pi^{2/3}}\frac{1}{d_{\rm L}}M_c^{{5}/{6}}f^{-7/6}\frac{1+\cos^2_\iota}{2}(-\beta)\left[1-\frac{1}{2}b\eta^{2/5}
(\pi \mathcal{M}_c f)^{-2/3}\right]\kappa^{-1/2}\mathcal{G}^{2/3} e^{i\Psi_+(f)},
\ena
where the phase is given by
\bea \Psi_+(f)&=&-2\psi_c+2\pi
ft_c-\frac{\pi}{4}+\frac{3}{128}(\pi\mathcal{M}_c f)^{-5/3}\left[1-\frac{4}{7}b\eta^{2/5}(\pi\mathcal{M}_c f)^{-2/3}\right],
\ena
where $\psi_c$  \textcolor{black}{ is} the phase of binary system at time $t_c$. The expression of the phase is consistent with that  \textcolor{black}{given} in \cite{will1994}.

 \textcolor{black}{Following a similar procedure, we can derive the Fourier components for the ``cross" and ``breathing" modes}, which are given by
\bea
\tilde{h}_{\times}(f)&=&\sqrt{\frac{5}{24}}\frac{1}{\pi^{2/3}}\frac{1}{d_{\rm L}}M_c^{5/6}f^{-7/6}\cos\iota(-\beta)\left[1-\frac{1}{2}b\eta^{2/5}
(\pi \mathcal{M}_c f)^{-2/3}\right]\kappa^{-1/2}\mathcal{G}^{2/3} e^{i\Psi_{\times}(f)}, \\
\tilde{h}_{b1}(f)&=&\sqrt{\frac{5}{24}}\frac{1}{\pi^{2/3}}\frac{1}{d_{\rm L}}M_c^{5/6}f^{-7/6}\sin^2\iota(-\xi\Gamma/4)\left[1-\frac{1}{2}b\eta^{2/5}
(\pi \mathcal{M}_c f)^{-2/3}\right]\kappa^{-1/2}\mathcal{G}^{4/3} e^{i\Psi_+(f)},\\
\tilde{h}_{b3}(f)&=&\sqrt{\frac{5}{48}}\frac{1}{\pi^{2/3}}\frac{1}{d_{\rm L}}M_c^{5/6}(2f)^{-7/6}\sin\iota(2\pi m f)^{-1/3}\kappa^{-1/2}\mathcal{G}\xi\mathcal{S}\left[1-\frac{1}{2}b\eta^{2/5}
(2\pi \mathcal{M}_c f)^{-2/3}\right]e^{i\Psi_{b3}(f)},
\ena
where the phases are
\bea
\Psi_{\times}(f)&=&\Psi_+(f)+\frac{\pi}{2}, \\
\Psi_{b3}(f)&=&-\psi_c+2\pi
ft_c-\frac{\pi}{4}+\frac{3}{256}(2\pi\mathcal{M}_c f)^{-5/3}\left[1-\frac{4}{7}b\eta^{2/5}(2\pi\mathcal{M}_c f)^{-2/3}\right].
\ena
Let us turn to  \textcolor{black}{the} $h_{b2}$ component. From Eq. (\ref{hb2}), we know that the phase is zero, and the value of $h_{b2}(t)$ depending on time $t$ is only through slowly  \textcolor{black}{varying function}  $f_s(t)$. So, the Fourier component $\tilde{h}_{b2}(f)$ is negligible in comparison with the other terms.

In order to  extend these results  \textcolor{black}{easily  to  high } post-Newtonian orders, we  \textcolor{black}{ rewrite the expressions of $h_{+}$ and $h_{\times}$ in the forms},
\bea\label{eq56} h_{+}(t)= \frac{2\beta\eta m x}{d_{\rm L}}H^{(0)}_{+},~~~
h_{\times}(t)=\frac{2\beta\eta m x}{d_{\rm L}}H^{(0)}_{\times},
\ena
where
\bea\label{eq57}
x=(2\pi mf_s)^{2/3},~~~
H_{+}^{(0)}=-(1+\cos^2\iota)\cos2\Phi(t),~~~
H_{\times}^{(0)}=-2\cos\iota\sin2\Phi(t).
\ena

A detector measures only a certain linear combination of the  \textcolor{black}{GW} components, called the response $h(t)$. For BD gravity, it is given by
\bea h(t)=F_{+}(\theta,\phi,\psi)h_{+}(t)+F_{\times}(\theta,\phi,\psi)h_{\times}(t)+F_{b}(\theta,\phi,\psi)h_{b}(t),
\ena
where $F_{+}$, $F_{\times}$ and $F_{b}$ are the detector antenna pattern functions, $\psi$ is the polarization angle as mentioned above, $(\theta,\phi)$ are angles describing the location of source on the sky, relative to the detector. In general these angles are
time-dependent. In the case of Einstein Telescope, considered in this paper, compact binary systems can be in band for hours, but almost all of the signal-to-noise ratio will be accumulated  \textcolor{black}{only at} the final minutes of the inspiral process. In the sequel, $(\theta,\phi,\psi)$ will be considered  \textcolor{black}{as constants } \footnote{ Note that with LISA, Doppler modulation due to the orbital motion, as well as spin precession, will allow for accurate determination of the angular parameters (see, for instance, \cite{triassintes} and references therein), but this is unlikely
to happen for BNS (or NSBH) signals in ET with Doppler modulation due to the Earth's rotation. Nevertheless, some improvement in parameter estimation can be expected, which for simplicity we do not take into account here.}.

The Fourier component of $h(t)$ becomes,
\bea
\tilde{h}(f)=F_+\tilde{h}_+(f)+F_{\times}\tilde{h}_{\times}(f)+F_b[\tilde{h}_{b1}(f)+\tilde{h}_{b3}(f)]\equiv\tilde{h}^{(1)}(f)+\tilde{h}^{(2)}(f),\ena
where
\bea\label{h1-f}
\tilde{h}^{(1)}(f)&=&\frac{M_c^{5/6}}{d_{\rm L}}\sqrt{\frac{5}{48}}\pi^{-2/3}(2f)^{-7/6}\left\{E(2\pi
mf)^{-1/3}+ES_{-1}(2\pi m f)^{-1}\right\} \nonumber
\\ &&\times\Theta(f_{\rm LSO}-f)\exp[i(2\pi ft_c-\pi/4+\psi(f))], \\
\label{h2-f}\tilde{h}^{(2)}(f)&=&2^{-1/2}\frac{M_c^{5/6}}{d_{\rm L}}\sqrt{\frac{5}{48}}\pi^{-2/3}f^{-7/6}\left\{[Qe^{-i\varphi_{(2,0)}}P_{(2,0)}+A]S_{-1}(\pi
mf)^{-2/3} +[Qe^{-i\varphi_{(2,0)}}P_{(2,0)}+A]\right\}\nonumber \\
&&\times\Theta(2f_{\rm LSO}-f)\exp[i(2\pi
ft_c-\pi/4+2\psi(f/2))], \ena
in which $\Theta(x)$ is the usual Heaviside function, $P_{(2,0)}$ and $\varphi_{(2,0)}$ are defined in Appendix B. The upper  \textcolor{black}{cutoff} frequency is dictated by the last stable orbit of the binary system, which marks the end of the inspiral regime and the onset of the finial merge. We assume that this occurs when the radiation frequency reaches $f=k f_{\rm LSO}$ for the \textcolor{black}{$k$-th} harmonic, with $f_{\rm LSO}=1/(6^{3/2}2\pi m)$ being the orbital frequency at the last stable orbit \footnote{\textcolor{black}{Note that, there is a small mistake in Eq. (41) in Ref. \cite{yunes2012}, where the coefficient of $-\frac{1}{4}\xi$ should be replaced by $-\frac{1}{2}\xi$. If taking into account this mistake, the formulae in (\ref{h1-f}) and (\ref{h2-f}) are consistent with the expressions of Eq. (55) and (54) in Ref. \cite{yunes2012}.}}. Note that, for the sources at cosmological distances, what enters the waveform is the {\emph{observed}} mass, which differs from the {\emph{physical}} mass by a factor $(1+z)$: $m_{\rm obs}=(1+z)m_{\rm phys}$ \cite{gw-book}. Throughout this paper, all the masses refer to the observed quantity if there is no special instruction. In these expressions, we have defined the following coefficients to  \textcolor{black}{characterize} the modifications of BD gravity,
\bea\label{other-definitions}
E=\kappa^{-1/2}\mathcal{G}\sin\iota F_b\xi\mathcal{S},~~~A=-\frac{1}{2}\xi\Gamma
\sin^2\iota F_b\kappa^{-1/2}\mathcal{G}^{4/3}, ~~
Q=(1-\frac{1}{2}\xi)\kappa^{-1/2}\mathcal{G}^{4/3},~~~
S_{-1}=-\frac{1}{2}b\kappa^{-2/5}\mathcal{G}^{8/15}.\ena
The uniform phase function \textcolor{black}{$\psi(f)$} is given by
\bea\label{varphi-f}
\psi(f)=-\psi_c+\frac{3}{256(2\pi\mathcal{M}_c f)^{5/3}}\sum_{i=-2}^{0}\psi_i(2\pi
mf)^{i/3}\ena with \bea \label{psi-2}
\psi_{-2}=-\frac{4}{7}b\kappa^{-2/5}\mathcal{G}^{8/15},~~~
\psi_{-1}=0,~~~ \psi_{0}=1.\ena
The quantity $Q$ describes the  \textcolor{black}{modifications} on the amplitudes of   \textcolor{black}{the} ``plus" and ``cross" modes. In addition, $A$, $E$, $S_{-1}$ and $Q$ together  \textcolor{black}{describe} the extra ``breathing" mode, which is absent in GR. The modification of the phase is described by $\psi_{-2}$. In the case with $E\rightarrow0$, $A\rightarrow0$, $S_{-1}\rightarrow0$, $\psi_{-2}\rightarrow0$, $Q\rightarrow1$ and $\mathcal{M}_c\rightarrow M_c$, the expression of $\tilde{h}(f)$  \textcolor{black}{reduces} to that  \textcolor{black}{of} GR. Expressions  \textcolor{black}{of} Eqs. (\ref{h1-f}), (\ref{h2-f}) and (\ref{varphi-f}) show that, in comparison with the waveform  \textcolor{black}{given}  in GR, the corrections of BD gravity are mainly at the low frequency range. Since the coefficients $E$, $S_{-1}$ and $\psi_{-2}$ all directly depend on \textcolor{black}{ the difference in sensitivities $\mathcal{S}$}, the corrections caused by the related terms \textcolor{black}{vanish for } the binary neutron star systems (if assuming the sensitivities of neutron stars are \textcolor{black}{the} same), \textcolor{black}{ as well as for binary black hole systems}. In addition, for the binary black hole systems with $s_i=0.5$, we have $\Gamma=0$, and the coefficient $A$ and its related terms  \textcolor{black}{ also vanish}. So, in comparison with GR, the difference of \textcolor{black}{the gravitational waveforms from} BD gravity is very small. For these reasons, in this paper, we shall only use the compact binary systems \textcolor{black}{ that are} composed of a neutron star and a black hole, in order to constrain BD gravity.

\subsection{Extension to high post-Newtonian orders}

In \textcolor{black}{GR and alternative theories} of gravity, the gravitational waveforms should include \textcolor{black}{high-order} PN terms to construct the real templates for the \textcolor{black}{GW} detectors. In the PN \textcolor{black}{approximations of GR, the waveforms} are expressed as expansions in \textcolor{black}{terms of } the orbital velocity $v$, \textcolor{black}{and have} been developed by many authors (see \cite{pn-review} and references therein). For the non-spinning compact objects, the best waveforms currently available are of 2.5 PN order in amplitude \cite{amplitude} and 3.5 PN order in phase \cite{phase}. In the scalar-tensor gravity, the equations of motion for non-spinning compact objects have been developed to 2.5 PN order \cite{bd-higher}, and the tensor and scalar gravitational waveforms have also been calculated up to 2 and 1.5 PN order, respectively \cite{lang2013,lang2014}.

In this subsection, we shall extend the waveforms and their Fourier components in BD gravity derived above to \textcolor{black}{high-order PN approximations}, in which the waveforms are linear combinations of harmonics in the orbital phase, and
\textcolor{black}{ the $k$-th harmonics is cutoff at $kf_{\rm LSO}$ in the frequency domain}. Including the higher PN orders in \textcolor{black}{waveforms}, in particular \textcolor{black}{ in  terms of orbital  phase, the corrections} could significantly alter
the signal-to-noise ratio of the GW sources \cite{chris}. Different from the previous works \cite{lang2013,lang2014}, we consider only the leading order corrections of waveforms (including polarization mode, amplitude and phase) caused by BD gravity in comparison with GR, and add these correction terms to the high PN \textcolor{black}{waveforms} of GR. Since the BD parameter $\omega_{\rm BD}$ \textcolor{black}{has been tightly constrained} by various experiments, the correction terms of \textcolor{black}{the gravitational waveforms} in BD gravity \textcolor{black}{are} expected to be very \textcolor{black}{small. And the corrections from  higher PN orders are expected to be even smaller than the leading-order ones}.

In the stationary phase approximation, the amplitude-corrected waveforms in GR is explicitly presented in \cite{chris}, in which the total Fourier component $\tilde{h}(f)$ is the sum of seven harmonics, i.e.
\bea\label{hf-total}
\tilde{h}(f)=\sum_{k=1}^{7}\tilde{h}^{(k)}(f).
\ena
\textcolor{black}{Taking} into account the corrections caused by BD gravity in comparison with GR, the expressions of harmonics $\tilde{h}^{(k)}(f)$ are revised to the following forms,
\bea
\tilde{h}^{(1)}(f)&=&\frac{M_c^{5/6}}{d_{\rm L}}\sqrt{\frac{5}{48}}\pi^{-2/3}(2f)^{-7/6}\left\{ES_{-1}(2\pi
m f)^{-1}+E(2\pi
mf)^{-1/3} \right. \nonumber \\
&&+e^{-i\varphi_{(1,1/2)}}P_{(1,1/2)}(2\pi
mf)^{1/3} \nonumber \\
&&+\left[e^{-i\varphi_{(1,3/2)}}P_{(1,3/2)}+e^{-i\varphi_{(1,1/2)}}P_{(1,1/2)}S_1\right](2\pi
mf) \nonumber \\
&&+\left[e^{-i\varphi_{(1,2)}}P_{(1,2)}+e^{-i\varphi_{(1,1/2)}}P_{(1,1/2)}S_{3/2}\right](2\pi
mf)^{4/3} \nonumber \\
&&\left.+\left[e^{-i\varphi_{(1,5/2)}}P_{(1,5/2)}+e^{-i\varphi_{(1,3/2)}}P_{(1,3/2)}S_{1}+e^{-i\varphi_{(1,1/2)}}P_{(1,1/2)}S_{2}\right](2\pi
mf)^{5/3}\right\} \nonumber\\
&&\times\Theta(f_{\rm LSO}-f)\exp[i(2\pi ft_c-\pi/4+\psi(f))], \nonumber\\
\tilde{h}^{(2)}(f)&=&2^{-1/2}\frac{M_c^{5/6}}{d_{\rm L}}\sqrt{\frac{5}{48}}\pi^{-2/3}f^{-7/6}\left\{\left[A+Qe^{-i\varphi_{(2,0)}}\right]S_{-1}(\pi
mf)^{-2/3}
\right. \nonumber \\ &&+\left[A+Qe^{-i\varphi_{(2,0)}}P_{(2,0)}\right] \nonumber \\
&&+\left[e^{-i\varphi_{(2,1)}}P_{(2,1)}+e^{-i\varphi_{(2,0)}}P_{(2,0)}S_1\right](\pi
mf)^{2/3} \nonumber \\
&&+\left[e^{-i\varphi_{(2,3/2)}}P_{(2,3/2)}+e^{-i\varphi_{(2,0)}}P_{(2,0)}S_{3/2}\right](\pi
mf) \nonumber \\
&&+\left[e^{-i\varphi_{(2,2)}}P_{(2,2)}+e^{-i\varphi_{(2,1)}}P_{(2,1)}S_{1}+e^{-\varphi_{(2,0)}}P_{(2,0)S_2}\right](\pi
mf)^{4/3} \nonumber \\
&&\left.+\left[e^{-i\varphi_{(2,5/2)}}P_{(2,5/2)}+e^{-i\varphi_{(2,3/2)}}P_{(2,3/2)}S_{1}+e^{-i\varphi_{(2,1)}}P_{(2,1)}S_{3/2}+e^{-i\varphi_{(2,0)}}P_{(2,0)}S_{5/2}\right](\pi
mf)^{5/3}\right\} \nonumber\\
&&\times\Theta(2f_{\rm LSO}-f)\exp[i(2\pi ft_c-\pi/4+2\psi(f/2))],\nonumber\\
\tilde{h}^{(3)}(f)&=&3^{-1/2}\frac{M_c^{5/6}}{d_{\rm L}}\sqrt{\frac{5}{48}}\pi^{-2/3}(2f/3)^{-7/6}\left\{e^{-i\varphi_{(3,1/2)}}P_{(3,1/2)}(2\pi mf/3)^{1/3}\right. \nonumber \\
&&+\left[e^{-i\varphi_{(3,3/2)}}P_{(3,3/2)}+e^{-i\varphi_{(3,1/2)}}P_{(3,1/2)}S_1\right](2\pi
mf/3) \nonumber \\
&&+\left[e^{-i\varphi_{(3,2)}}P_{(3,2)}+e^{-i\varphi_{(3,1/2)}}P_{(3,1/2)}S_{3/2}\right](2\pi
mf/3)^{4/3} \nonumber \\
&&\left.+\left[e^{-i\varphi_{(3,3/2)}}P_{(3,3/2)}S_1+e^{-i\varphi_{(3,1/2)}}P_{(3,1/2)}S_{2}\right](2\pi
mf/3)^{5/3} \right\} \nonumber\\
&&\times\Theta(3f_{\rm LSO}-f)\exp[i(2\pi ft_c-\pi/4+3\psi(f/3))],\nonumber\\
\tilde{h}^{(4)}(f)&=&4^{-1/2}\frac{M_c^{5/6}}{d_{\rm L}}\sqrt{\frac{5}{48}}\pi^{-2/3}(f/2)^{-7/6}\left\{e^{-i\varphi_{(4,1)}}P_{(4,1)}(\pi mf/2)^{2/3}\right. \nonumber \\
&&+\left[e^{-i\varphi_{(4,2)}}P_{(4,2)}+e^{-i\varphi_{(4,1)}}P_{(4,1)}S_1\right](\pi
mf/2)^{4/3} \nonumber \\
&&\left.+\left[e^{-i\varphi_{(4,5/2)}}P_{(4,5/2)}+e^{-i\varphi_{(4,1)}}P_{(4,1)}S_{3/2}\right](\pi
mf/2)^{5/3} \right\} \nonumber\\
&&\times\Theta(4f_{\rm LSO}-f)\exp[i(2\pi ft_c-\pi/4+4\psi(f/4))],\nonumber\\
\tilde{h}^{(5)}(f)&=&5^{-1/2}\frac{M_c^{5/6}}{d_{\rm L}}\sqrt{\frac{5}{48}}\pi^{-2/3}(2f/5)^{-7/6}\left\{e^{-i\varphi_{(5,3/2)}}P_{(5,3/2)}(2\pi mf/5)\right. \nonumber \\
&&\left.+\left[e^{-i\varphi_{(5,5/2)}}P_{(5,5/2)}+e^{-i\varphi_{(5,3/2)}}P_{(5,3/2)}S_{1}\right](2\pi
mf/5)^{5/3} \right\} \nonumber\\
&&\times\Theta(5f_{\rm LSO}-f)\exp[i(2\pi ft_c-\pi/4+5\psi(f/5))],\nonumber\\
\tilde{h}^{(6)}(f)&=&6^{-1/2}\frac{M_c^{5/6}}{d_{\rm L}}\sqrt{\frac{5}{48}}\pi^{-2/3}(f/3)^{-7/6}e^{-i\varphi_{(6,2)}}P_{(6,2)}(\pi mf/3)^{4/3} \nonumber \\
 &&\times\Theta(6f_{\rm LSO}-f)\exp[i(2\pi ft_c-\pi/4+6\psi(f/6))],\nonumber\\
\tilde{h}^{(7)}(f)&=&7^{-1/2}\frac{M_c^{5/6}}{d_{\rm L}}\sqrt{\frac{5}{48}}\pi^{-2/3}(2f/7)^{-7/6}e^{-i\varphi_{(7,5/2)}}P_{(7,5/2)}(2\pi mf/7)^{5/3} \nonumber \\
 &&\times\Theta(7f_{\rm LSO}-f)\exp[i(2\pi ft_c-\pi/4+7\psi(f/7))],\nonumber\ena
where $P_{(m,n)}$, $\varphi_{(m,n)}$, $S_{i}~(i\ge 1)$ are all given in Appendix B. The other parameters, including $E$, $A$, $Q$, $S_{-1}$, are defined by Eq. (\ref{other-definitions}). In these expressions, the phase function $\psi(f)$ is given by
\bea\label{full-psi}
\psi(f)=-\psi_c+\frac{3}{256(2\pi\mathcal{M}_c f)^{5/3}}\sum_{i=-2}^{7}\psi_i(2\pi
mf)^{i/3},
\ena
where $\psi_{-2}$ and $\psi_{-1}$ are given by Eq. (\ref{psi-2}), and $\psi_{i}~(i\ge 0)$ are given in Appendix B.

\begin{figure}
\begin{center}
\centerline{\includegraphics[width=15cm]{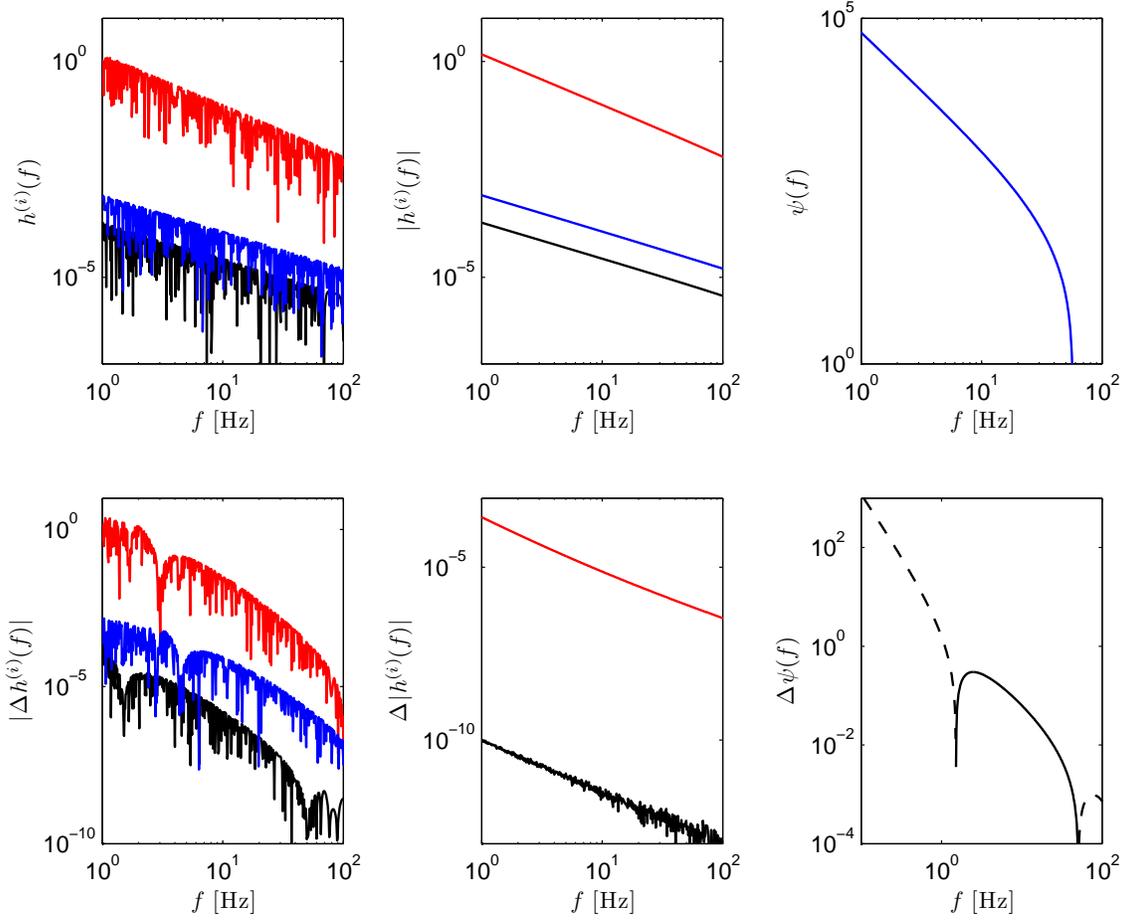}}
\end{center}\caption{Upper panels show the amplitudes and phase of the Fourier components $\tilde{h}^{(i)}(f)$ in  \textcolor{black}{BD} gravity.
Lower panels show $|\tilde{h}^{(i)}_{\rm BD}-\tilde{h}^{(i)}_{\rm GR}|$ (left), $||\tilde{h}^{(i)}_{\rm BD}|-|\tilde{h}^{(i)}_{\rm GR}||$ (middle) and $(\psi_{\rm BD}-\psi_{\rm GR})$ (right). In  \textcolor{black}{the} left and middle panels,  \textcolor{black}{the} black lines denote the results of  \textcolor{black}{the} first harmonic with $i=1$,  \textcolor{black}{the} red lines are those with $i=2$, and  \textcolor{black}{the} blue lines are those with $i=3$. In  \textcolor{black}{the right panels}, the negative values of the function are depicted by  \textcolor{black}{the} broken line. In this figure we have adopted the model with  \textcolor{black}{the parameters chosen as}, $m_1=10M_{\odot}$, $m_2=1.4M_{\odot}$, $\theta=\phi=\psi=\iota=0$, $\xi=0.001$, $d_{\rm L}=10^3$Mpc. Note that the units of  \textcolor{black}{the} vertical axis in  \textcolor{black}{the} left and middle panels are all  \textcolor{black}{rescaled by a factor $10^{-20}{\rm Hz}^{-1}$}. }\label{f2}
\end{figure}

From the expression of $\tilde{h}(f)$ we find that the corrections caused by BD gravity exist both in the amplitudes $\tilde{h}^{i}(f)~(i=1,2)$ and the phase $\psi(f)$. In order to investigate which effect is dominant \textcolor{black}{for} a typical binary system, we plot the waveforms $\tilde{h}^{i}(f)~(i=1,2,3)$ and the difference between GR and BD gravity in Fig. \ref{f2}. In this system, we choose the mass of \textcolor{black}{the} black hole \textcolor{black}{as} $m_1=10M_{\odot}$ with \textcolor{black}{the} sensitivity $s_1=0.5$, the mass of \textcolor{black}{the} neutron star \textcolor{black}{as} $m_2=1.4M_{\odot}$ with \textcolor{black}{the} sensitivity $s_2=0.2$, and the BD parameter \textcolor{black}{as} $\xi=0.001$. \textcolor{black}{Note that, this number of $\xi$ has already been ruled out by the Cassini experiment \cite{bound}, which was used here only for an illustrative purpose.}
 The left panels show that the second harmonic is much larger than the other ones, which dominates the signal-to-noise ratio of the event. \textcolor{black}{The middle} panels show that the values of $|\tilde{h}^{(i)}_{\rm BD}-\tilde{h}^{(i)}_{\rm GR}|$ are comparable \textcolor{black}{to} those of $\tilde{h}^{(i)}_{\rm BD}$ or $\tilde{h}^{(i)}_{\rm GR}$ for any \textcolor{black}{given} frequency $f$, which indices that the correction effects are significant in the BD gravity with $\xi=0.001$. However, if ignoring the correction effects in the phase terms and considering only the amplitudes of the waveforms, we find that the values of $||\tilde{h}^{(i)}_{\rm BD}|-|\tilde{h}^{(i)}_{\rm GR}||$ become much smaller than those of $|\tilde{h}^{(i)}_{\rm BD}|$ or $|\tilde{h}^{(i)}_{\rm GR}|$. So, we conclude that the dominant effects of BD gravity are caused by the modification in the phase  \textcolor{black}{terms, rather than  in } the amplitude terms, which is consistent with the arguments  \textcolor{black}{given}  in \cite{will1994,will-lisa}. From the expression of $\psi(f)$ in Eq. (\ref{full-psi}), we observe that the modification on the phase  \textcolor{black}{terms} has two effects: The chirp mass $M_c$ in the denominator is replaced by $\mathcal{M}_c$, and an extra term $\psi_{-2}(2\pi mf)^{-2/3}$. Compared with the phase  \textcolor{black}{given} in GR, the first effect increases the value of $\psi(f)$, and the  \textcolor{black}{latter} decreases it. From the right panel of Fig. \ref{f2}, we find  \textcolor{black}{that, in the low} frequency range $f<1.5$Hz, the first effect is dominant, and  \textcolor{black}{ in the high} frequency range $f>1.5$Hz, \textcolor{black}{ the latter}   is dominant.

\section{Testing BD gravity using Einstein Telescope \label{section3}}

\subsection{Einstein Telescope and the estimation of GW parameters\label{section3.1}}

The \textcolor{black}{gravitational waveforms} depend not only on the parameters of the binary system, but also on the parameters  \textcolor{black}{of the theory concerned}  (\textcolor{black}{For example, in BD} gravity, it is $\omega_{\rm BD}$). By the matched-filter analysis of  \textcolor{black}{GW} observations, one can determine all the parameters together. In this paper, we shall focus on the observation of GW sources by ET, a third-generation ground-based  \textcolor{black}{GW} detector. Although the basic design of ET is still under  \textcolor{black}{discussions, one} possibility is to have three interferometers with $60^{\circ}$ opening angles  \textcolor{black}{and} 10 km arm lengths, arranged in an equilateral triangle \cite{et}. The corresponding antenna pattern functions of ET for different polarization modes of GWs are given in Appendix C. The scientific potentials of ET have been studied by many authors \cite{sayth,zhao2011,gair2,cai,zhu,mock,gair,et-science,d1,d2,d3,vilta,zhu2017}.

The performance of a  \textcolor{black}{GW} detector is characterized by the one-side noise {\emph{power spectral density}} $S_h(f)$ (PSD), which plays an important role in the signal analysis. We take the noise PSD of  \textcolor{black}{ET} to be \cite{freise,zhao2011}
\bea
S_h(f)=S_0\left[x^{p_1}+a_1x^{p_2}+a_2\frac{1+b_1x+b_2x^2+b_3x^3+b_4x^4+b_5x^5+b_6x^6}{1+c_1x+c_2x^2+c_3x^3+c_4x^4}\right],
\label{et-S} \ena where $x\equiv f/f_0$ with $f_0=200$Hz, and
$S_0=1.449\times 10^{-52}\,{\rm Hz}^{-1}$. The other parameters
are as follows: \bea
p_1=-4.05,~~p_2=-0.69,&&
a_1=185.62,~~a_2=232.56,\nonumber\\
b_1=31.18,~~b_2=-64.72,~~b_3=52.24,&&b_4=-42.16, ~~b_5=10.17, ~~b_6=11.53\nonumber\\
c_1=13.58,~~c_2=-36.46,&&c_3=18.56, ~~c_4=27.43. \ena For  \textcolor{black}{ the purpose of  data
analysis}, the noise PSD is assumed to be essentially
infinite below a certain  \textcolor{black}{low} cutoff frequency $f_{\rm lower}$
(see the review \cite{gwreview}). For ET we take this to be
$f_{\rm lower}=1~{\rm Hz}$.

For any given binary system, the waveforms in Eq.~(\ref{hf-total}) depend on nine system
parameters $({M}_c,\eta,t_c,\psi_c,\iota,\theta,\phi,\psi,d_{\rm L})$ and one gravity parameter $\xi$ (or $\omega_{\rm BD}$). By maximizing the correlation between a template waveform that depends on a set of parameters $p_i$ ($i=1,2,3,\cdot\cdot\cdot$) and a measured signal, the matched filtering provides a natural way to estimate the parameters of the signal and their errors. With a given detector noise $S_h(f)$, we employ the Fisher
matrix approach \cite{gwfisher}. Comparing with the Markov chain
Monte Carlo (MCMC) analysis, the Fisher information matrix
analysis is simple and accurate enough to estimate the detection
abilities of the future experiments. In the case of a single
interferometer $A$ ($A = 1, 2, 3$ for ET), the Fisher matrix is given by \cite{will1994}
\bea \Lambda^{\rm A}_{ij}=\langle \tilde{h}^{\rm A}_i(f),
\tilde{h}^{\rm A}_j(f)\rangle,~~~\tilde{h}^{\rm A}_i(f)=\partial
\tilde{h}^{\rm A}(f)/\partial p_i, \ena where $\tilde{h}^{\rm A}(f)$
is the output of  \textcolor{black}{the} interferometer $A$, and $p_i$ denote the free
parameters to be estimated, which are \be
({M}_c,\eta,t_c,\psi_c,\cos\iota,\cos\theta,\phi,\psi,\ln{d_{\rm L}},\xi).
\ee Note that, in this paper, we fix the sensitivities as follows: For neutron  \textcolor{black}{stars} $s_2=0.2$, and for black  \textcolor{black}{holes} $s_1=0.5$.
The angular brackets denote the scalar product, which, for any
two  \textcolor{black}{given functions $a(t)$ and $b(t)$,} is defined as \bea \langle
a,b\rangle=4\int_{f_{\rm lower}}^{f_{\rm upper}}\frac{\tilde{a}(f)
\tilde{b}^*(f)+\tilde{a}^*(f)\tilde{b}(f)}{2}\frac{df}{S_h(f)},
\label{innerproduct} \ena where $\tilde{a}$ and $\tilde{b}$ are
the Fourier transforms of the functions $a(t)$ and $b(t)$. The
Fisher matrix for the combination of the three independent
interferometers is then \be \label{lambda-definition}\Lambda_{ij} = \sum_{{\rm A}=1}^3
\Lambda^{\rm A}_{ij}. \ee
Once the total Fisher matrix $\Lambda_{ij}$ is derived, an estimate of rms error, $\Delta p_{i}$, in measuring the parameter $p_i$ can then be calculated in the limit of large signal-to-noise ratio, by taking the square root of the diagonal elements of the inverse of  \textcolor{black}{the} Fisher matrix,
\bea\label{delta-definition}
\Delta p_i=(\Sigma_{ii})^{1/2},~~~\Sigma=\Lambda^{-1}.
\ena
The correlation coefficients between parameters $p_i$ and $p_j$ are given by
\bea
c_{ij}=\Sigma_{ij}/(\Sigma_{ii}\Sigma_{jj})^{1/2}.
\ena
Note that, in the limit  \textcolor{black}{case}  $c_{ij}=0$ for any $i\neq j$, the error becomes $\Delta p_i \rightarrow(\Lambda_{ii})^{-1/2}$, which is also equivalent to the case in which all the other parameters, but $p_{i}$, are fixed at the parameter estimation.

\begin{figure}
\begin{center}
\centerline{\includegraphics[width=10cm]{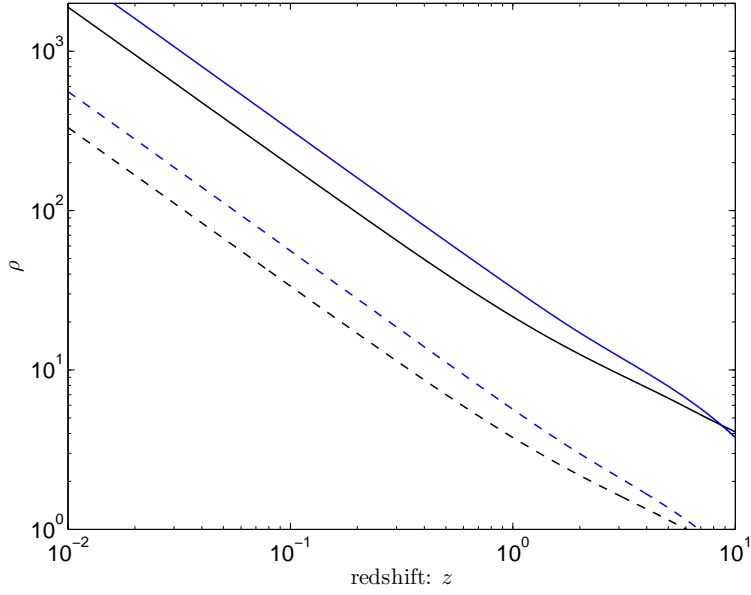}}
\end{center}\caption{Signal-to-noise ratio for the sources at different  \textcolor{black}{redshifts}.  \textcolor{black}{The black} solid line shows the result with  \textcolor{black}{the} parameters $m_{1,{\rm phys}}=2M_{\odot}$, $\theta=\phi=\psi=\iota=0$;
 \textcolor{black}{the } blue solid line shows  \textcolor{black}{the results } with $m_{1,{\rm phys}}=10M_{\odot}$, $\theta=\phi=\psi=\iota=0$.  \textcolor{black}{The black} dashed line shows the result with  \textcolor{black}{the} parameters $m_{1,{\rm phys}}=2M_{\odot}$, $\theta=\phi=\psi=\iota=\pi/2$, and blue dashed line shows that with $m_{1,{\rm phys}}=10M_{\odot}$, $\theta=\phi=\psi=\iota=\pi/2$.
For the other parameters, in all cases we have adopted the same  \textcolor{black}{values:} $t_c=0$, $\psi_c=0$, $\xi=0.001$, $m_{2,{\rm phys}}=1.4M_{\odot}$.}\label{f6}
\end{figure}

The inner product also allows us to write the signal-to-noise
ratios $\rho^{\rm A}~(A=1,2,3)$ in a compact way: \be \rho^{\rm A} =
\sqrt{\langle \tilde{h}^{\rm A}(f), \tilde{h}^{\rm A}(f) \rangle}. \ee
The combined signal-to-noise ratio for the network of the three
independent interferometers is then \be \label{rho-definition}\rho = \left[ \sum_{{\rm A} =
1}^3 \left(\rho^{\rm A}\right)^2 \right]^{1/2}. \ee
In Fig. \ref{f6}, we plot the signal-to-noise ratio $\rho$ for different compact binary systems at different positions,  \textcolor{black}{in which} we find that the value of $\rho$ strongly depends on the mass, the redshift, as well as their positions in the sky. For the given position, orbital and polarization angles, the higher redshift and/or  \textcolor{black}{the  larger mass of the black hole} follow the lower signal-to-noise ratio. If the sources are at the optimum  \textcolor{black}{position,  $\theta=\psi=\iota=0$, and the mass of the black hole is $m_{1,{\rm phys}}=2M_{\odot}$, we have $\rho>8$ for $z<3.82$. If } $m_{1,{\rm phys}}=10M_{\odot}$, it becomes $\rho>8$ so long as $z<4.93$. In both cases, we find ET could detect the binary systems at very high  \textcolor{black}{redshifts}.
On the contrary, if the sources are at the  \textcolor{black}{position $\theta=\psi=\iota=\pi/2$}, we have $\rho>8$  \textcolor{black}{for $z<0.43$ and} $m_{1,{\rm phys}}=2M_{\odot}$, and $\rho>8$  \textcolor{black}{for $z<0.70$ and} $m_{1,{\rm phys}}=10M_{\odot}$. In addition,  \textcolor{black}{from numerical  calculations}, we find that the signal-to-noise ratio $\rho$ is independent of the position angle $\phi$, which is determined by the equilateral triangle structure of  \textcolor{black}{ET}.

\begin{figure}
\begin{center}
\centerline{\includegraphics[width=10cm]{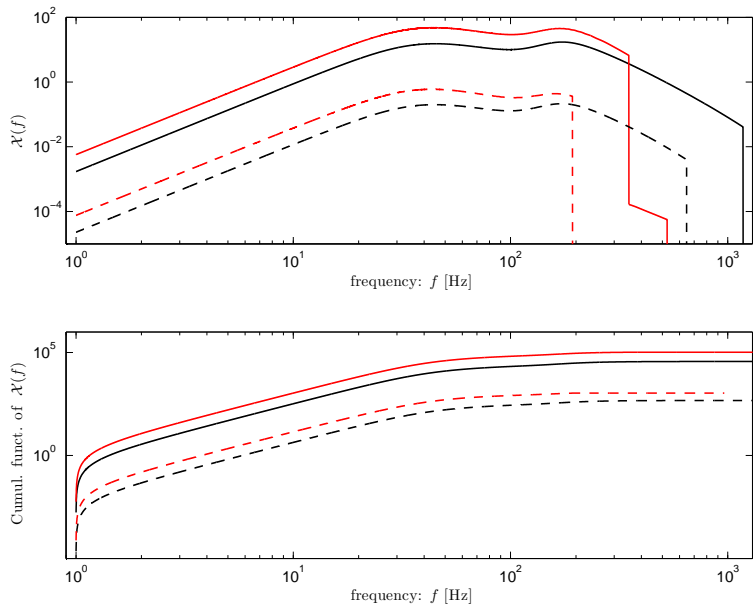}}
\end{center}\caption{The function $\mathcal{X}(f)$ for different  \textcolor{black}{frequencies} $f$ (upper panel), and its cumulative function from $f_{\rm lower}$ to $f$ (lower panel).  \textcolor{black}{The black} solid line shows the result with $m_{1,{\rm phys}}=2M_{\odot}$ and $z=0.1$, and  \textcolor{black}{the} red solid line shows the result with $m_{1,{\rm phys}}=10M_{\odot}$ and $z=0.1$.  \textcolor{black}{The black} dashed line shows the result with $m_{1,{\rm phys}}=2M_{\odot}$ and $z=1$, and
 \textcolor{black}{the}  red dashed line shows the result with $m_{1,{\rm phys}}=10M_{\odot}$ and $z=1$. For the other parameters, in both cases we have  \textcolor{black}{set} $\theta=\phi=\psi=\iota=0$, $t_c=0$, $\psi_c=0$, $\xi=0.001$, $m_{2,{\rm phys}}=1.4M_{\odot}$.}\label{f7}
\end{figure}

In order to study the contribution of signal at each frequency band to the total signal-to-noise ratio $\rho$, we define the following quantity,
\bea
\mathcal{X}(f)\equiv \sum_{{\rm A}=1}^{3} \frac{4f(\Delta \ln f)|\tilde{h}^{\rm A}(f)|^2}{S_n(f)},
\ena
where we bin the frequency band with $\Delta\ln (f/{\rm Hz})=0.001$ in this paper. It is obvious that $\mathcal{X}(f)$ at each frequency  \textcolor{black}{describes} the relative contribution of the signal-to-noise ratio at this single frequency band $f$, and the total signal-to-noise ratio $\rho^2$ defined in Eq. (\ref{rho-definition}) is the cumulative function of $\mathcal{X}(f)$ from $f=f_{\rm lower}$ to $f=f_{\rm upper}$. In Fig. \ref{f7} we plot the function $\mathcal{X}(f)$ and  \textcolor{black}{ its} cumulative function from $f_{\rm lower}$ to $f$ for different objects. All the plots clearly show that the  \textcolor{black}{major} contribution to the total $\rho$ comes from the signal at  \textcolor{black}{the} frequency range $f\in (30,300)$Hz, which is caused by the fact that the noise PSD of ET  \textcolor{black}{is minimized about} $f\sim 200$Hz.

\subsection{Potential constraint on the parameter $\omega_{\rm BD}$ by ET}

\subsubsection{ \textcolor{black}{Dependence} on the mass of black hole}

As mentioned above, the corrections of the \textcolor{black}{gravitational waveforms} in BD gravity strongly depends on \textcolor{black}{the difference in sensitivities $\mathcal{S}$},  \textcolor{black}{so} we expect that only the compact binary systems including a neutron star and a black hole can well constrain the parameter $\omega_{\rm BD}$. So, in this paper, we shall only focus on this kind of systems. For the neutron star, we assume  \textcolor{black}{its physical mass} is $m_{2,{\rm phys}}=1.4M_{\odot}$ and the sensitivity parameter  \textcolor{black}{is} $s_2=0.2$. For the black hole, we assume that  \textcolor{black}{its physical mass is  in the range} $m_{2,{\rm phys}}\in(2M_{\odot},100M_{\odot})$, the sensitivity parameter is $s_1=0.5$, and the spin is zero.

\begin{figure}
\begin{center}
\centerline{\includegraphics[width=10cm]{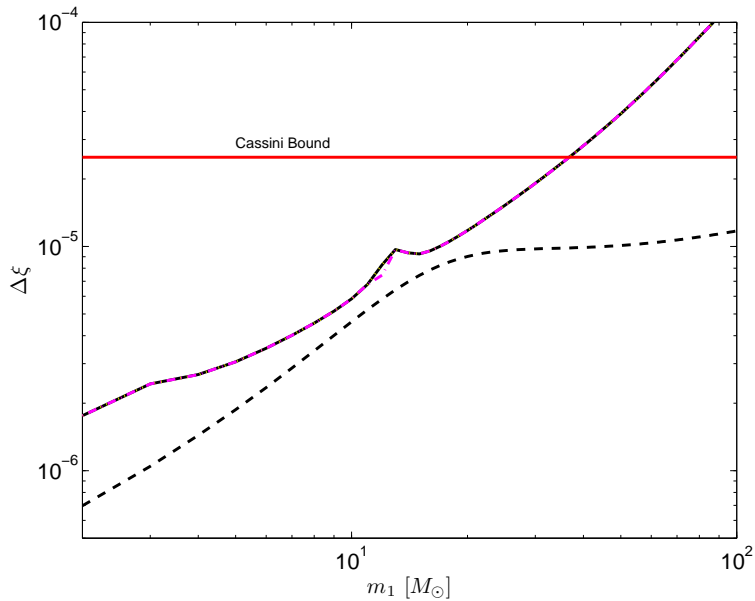}}
\end{center}\caption{ The dependence of rms error $\Delta\xi$ on the black hole mass $m_1$. The other parameters in  \textcolor{black}{the} fiducial model are  \textcolor{black}{set to}
$m_2=1.4M_{\odot}$, $t_c=0$, $\psi_c=0$, $\theta=\phi=\psi=\iota=0$, $\xi=0$, $d_{\rm L}=10^3$Mpc.  \textcolor{black}{The black} solid line shows the result in the case where all  \textcolor{black}{the} ten parameters are set free.  \textcolor{black}{The} black dashed line shows  \textcolor{black}{the result}  in the case where only $\xi$ is set free.  \textcolor{black}{The yellow} dashed line (which is overlapped with the black solid line) shows the result in the case where only  \textcolor{black}{the} parameters $(M_c,\eta,t_c,\psi_c,\psi,\iota,\xi,\ln d_{\rm L})$ are set free. In  \textcolor{black}{the} magenta dash-dotted line, we consider ten free parameters, but include only the phase correction in the  \textcolor{black}{gravitational waveforms}. \textcolor{black}{For comparison, we also plot the Cassini bound with red solid line.}
}\label{f0}
\end{figure}

Let us investigate what kind of systems can give a better  \textcolor{black}{constraint} on the  \textcolor{black}{BD parameter. Let us first fix the following model parameters as} $\theta=\phi=\psi=\iota=0$, $t_c=\psi_c=0$, $d_{\rm L}=10^3$Mpc, $m_2=1.4M_{\odot}$, and $\xi=0$ in the fiducial model.  \textcolor{black}{Then, we} study the effect of $m_1$ on the value of $\Delta\xi$. Here, since $z\ll 1$, we ignore the difference between the observed masses $m_{i}$ and the physical masses $m_{i,{\rm phys}}$. For each case, we solve the Fisher information matrix $\Lambda_{ij}$ with ten free parameters, and derive the quantity $\Delta\xi$ by using the relation  \textcolor{black}{given } in Eq. (\ref{delta-definition}), which  \textcolor{black}{is} plotted in Fig. \ref{f0},  \textcolor{black}{denoted by the black solid line}. This figure clearly shows that a smaller mass $m_1$ of  \textcolor{black}{the} black hole  \textcolor{black}{gives}  a lower value $\Delta\xi$, i.e. the more stringent constraint on  \textcolor{black}{the} parameter $\omega_{\rm BD}$  \textcolor{black}{is obtained}. For the binary system with $m_1=2M_{\odot}$, we have $\Delta\xi=1.76\times10^{-6}$, which is equivalent to the constraint $\omega_{\rm BD}>0.57\times10^6$. However, if $m_1=100M_{\odot}$, the error of $\xi$ becomes $\Delta\xi=1.27\times10^{-4}$, i.e. $\omega_{\rm BD}>0.79\times10^4$. As mentioned above, in this calculation, we have taken into account the correlation between $\xi$ and the other parameters in the analysis. If considering the limit case, in which only the parameter $\xi$ is set free,  \textcolor{black}{while all the other parameters are } fixed, then we calculate the errors $\Delta\xi=\Lambda_{\xi\xi}^{-1/2}$ for different $m_1$, which  \textcolor{black}{is} also plotted in Fig. \ref{f0},  \textcolor{black}{denoted by the black dashed line}. Comparing  \textcolor{black}{it} with the black solid line, we find that the results in these two cases are quite different, which shows that the cross-correlations between $\xi$ and other parameters can significantly weaken the constraints on $\xi$. However, in both cases, the tendencies between $\Delta\xi$ and $m_1$ are  \textcolor{black}{the} same, which are different from the results of signal-to-noise ratio $\rho$  \textcolor{black}{ploted}  in Fig. \ref{f6}.

\begin{figure}
\begin{center}
\centerline{\includegraphics[width=10cm]{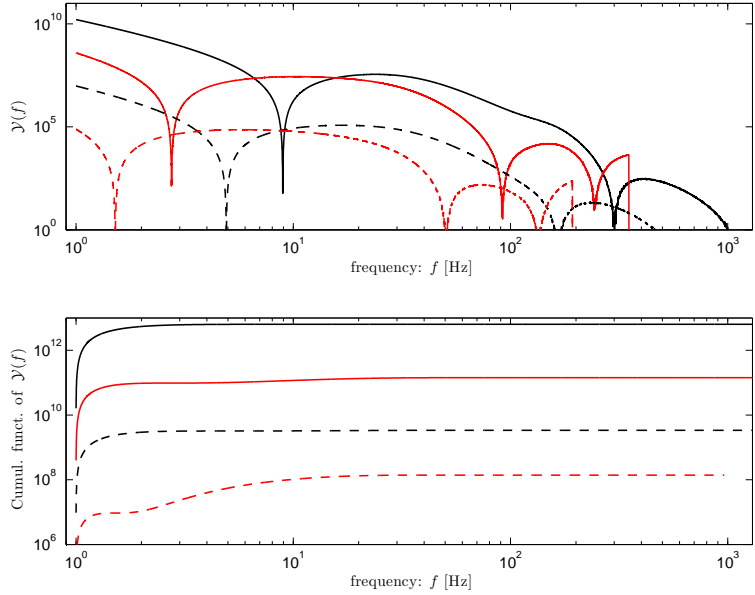}}
\end{center}\caption{The function $\mathcal{Y}(f)$ for different frequency $f$ (upper panel), and its cumulative function from $f_{\rm lower}$ to $f$ (lower panel).  \textcolor{black}{The black} solid line shows the result with $m_{1,{\rm phys}}=2M_{\odot}$ and $z=0.1$, and  \textcolor{black}{the} red solid line shows the result with $m_{1,{\rm phys}}=10M_{\odot}$ and $z=0.1$.  \textcolor{black}{The black} dashed line shows the result with $m_{1,{\rm phys}}=2M_{\odot}$ and $z=1$, and  \textcolor{black}{the}  red dashed line shows the result with $m_{1,{\rm phys}}=10M_{\odot}$ and $z=1$. For the other parameters, in  \textcolor{black}{all cases we have set} $\theta=\phi=\psi=\iota=0$, $m_{2,{\rm phys}}=1.4M_{\odot}$.}\label{f8}
\end{figure}

The Fisher matrix component $\Lambda_{\xi\xi}$  \textcolor{black}{describes} the sensitivity of  \textcolor{black}{ET} on the parameter $\xi$ (which is equivalent to $\omega_{\rm BD}$). In order to quantify the contribution of each frequency band to the total $\Lambda_{\xi\xi}$, we define the following quantity,
\bea
\mathcal{Y}(f)\equiv \sum_{{\rm A}=1}^{3} \frac{4f(\Delta \ln f)|\partial{\tilde{h}^{\rm A}}(f)/\partial\xi|^2}{S_n(f)}
\ena
where we bin the frequency band with $\Delta\ln (f/{\rm Hz})=0.001$ in this paper. The Fisher matrix component $\Lambda_{\xi\xi}$ defined in Eq. (\ref{lambda-definition}) is the cumulative function of $\mathcal{Y}(f)$ from $f=f_{\rm lower}$ to $f=f_{\rm upper}$. In Fig. \ref{f8} we plot the function $\mathcal{Y}(f)$ and  \textcolor{black}{ its} cumulative function from $f_{\rm lower}$ to $f$ for various objects. Different from the results  \textcolor{black}{given in Fig. \ref{f7}, } the plots clearly show that the main contribution to the total $\Lambda_{\xi\xi}$ comes from the signal at the lowest frequency range $f\sim f_{\rm lower}$, which is understandable since the waveform difference between BD gravity and GR is mainly at the low frequency range. If increasing the mass of  \textcolor{black}{the} black hole $m_1$, the frequency, in which the waveform difference is significant,  \textcolor{black}{will become} lower. Thus, in the sensitive frequency band $f>1$Hz of ET, the effect of BD gravity become  \textcolor{black}{weaker}, which explains  \textcolor{black}{why a large $m_1$ gives rise to a weaker constraint on the } parameter $\xi$.

In Section \ref{section2}, we know that the waveform correction caused by BD gravity can be divided into two parts: One is the correction in the phase term $\psi(f)$, and the other is  \textcolor{black}{in} the amplitudes of $\tilde{h}^{(1)}(f)$ and $\tilde{h}^{(2)}(f)$. In the previous works \cite{will1994,will-lisa}, the authors have only considered the correction in the phase term. Here, we will investigate how the corrections of  \textcolor{black}{the  amplitudes can} influence the value of $\Delta\xi$. In Fig. \ref{f0}, we plot the rms error $\Delta\xi$ ( \textcolor{black}{the} magenta line) in which only phase  \textcolor{black}{corrections are} considered. We find that the values of $\Delta\xi$ in this case is quite similar to those in the case including both  \textcolor{black}{ phase   and amplitude corrections}. So, we conclude that, the amplitude correction in the \textcolor{black}{gravitational waveforms} can only slightly influence the value $\Delta\xi$ at $m_{1}\sim 12M_{\odot}$.

\subsubsection{ \textcolor{black}{Dependance} on the confirmation of electromagnetic counterpart}

The coalescing binaries composed of a neutron star and a black hole could also cause the short-hard $\gamma$-ray bursts \cite{gamma-ray}. Many groups and telescopes tried to detect the electromagnetic counterparts of the GW bursts, by which one can determine the redshift of the burst. Combining the GW observation, which can determine the luminosity distance of the bursts independently, this kind of GW bursts can be treated as the standard sirens to study the expansion history of the universe \cite{schutz}. Here, we should mention that once the electromagnetic counterparts of the bursts are identified, their sky-positions are also confirmed. So, the uncertainties of the position parameters ($\theta,\phi$) should be excluded in the determination of  \textcolor{black}{the} parameter $\xi$. In order to investigate whether or not the value of $\Delta\xi$ can be significantly reduced for the sources with confirmed  \textcolor{black}{sky-positions}, we repeat the calculation with different black-hole  \textcolor{black}{masses}, and consider only eight free parameters ($M_c,\eta,\psi,\iota,t_c,\psi_c,d_{\rm L},\xi$), and plot the results of $\Delta\xi$ in Fig. \ref{f0} ( \textcolor{black}{the} yellow dotted line). We are surprised to find that the values of $\Delta\xi$ in this case are nearly  \textcolor{black}{the same as} those in the case with ten free parameters, which indices that cross-correlation between sky-position parameters and $\xi$ is weak. So, we conclude that the identification of electromagnetic counterparts of GW bursts cannot  \textcolor{black}{significantly} improve the constraint on the parameter $\xi$.

\subsubsection{ Dependance on the sky-position, inclination and polarization angles of the sources}

\begin{figure}
\begin{center}
\centerline{\includegraphics[width=15cm]{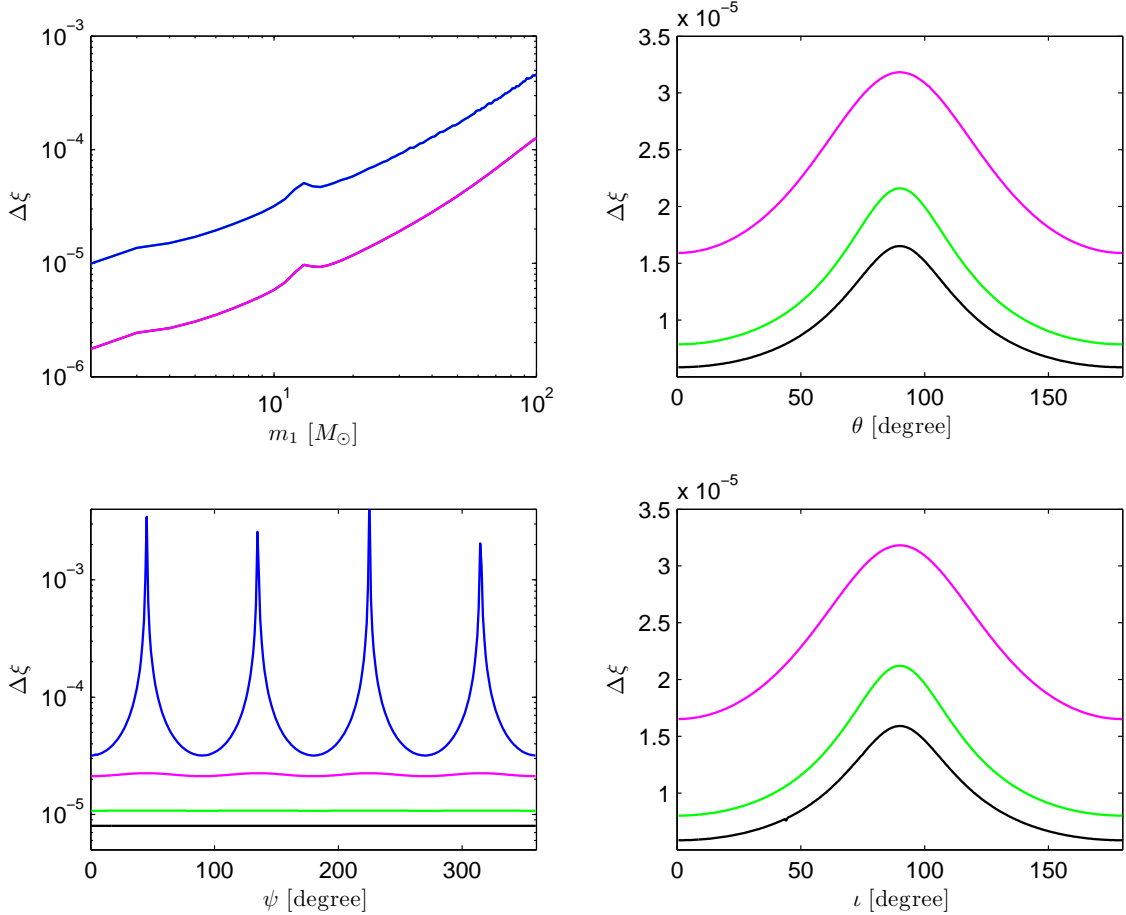}}
\end{center}\caption{ {\bf Upper left panel:}  \textcolor{black}{ The magenta} line shows the value of $\Delta\xi$ in the case with \textcolor{black}{the} fiducial model $\theta=\psi=\iota=0$, and  \textcolor{black}{the blue}
 line shows \textcolor{black}{the result} with $\theta=\psi=\iota=\pi/2$. In both cases, we have chosen $m_2=1.4M_{\odot}$, $\xi=0$, $d_{\rm L}=10^3$Mpc.\\
{\bf Upper right panel:}\textcolor{black}{The black}, green and magenta lines show the values of $\Delta\xi$ in the cases with \textcolor{black}{the} fiducial model $\iota=0$, $\iota=\pi/4$, $\iota=\pi/2$, respectively. In all cases, we have chosen $m_1=10M_{\odot}$, $m_2=1.4M_{\odot}$, $\xi=0$, $d_{\rm L}=10^3$Mpc, $\psi=0$. \\
{\bf Lower left panel:} \textcolor{black}{The black}, green, magenta and blue lines show the values of $\Delta\xi$ in the cases with \textcolor{black}{the} fiducial model $(\theta=\pi/4, \iota=0)$, $(\theta=\pi/4, \iota=\pi/4)$, $(\theta=\pi/4, \iota=\pi/2)$, $(\theta=\pi/2, \iota=\pi/2)$, respectively. In all cases, we have chosen $m_1=10M_{\odot}$, $m_2=1.4M_{\odot}$, $\xi=0$, $d_{\rm L}=10^3$Mpc.\\
{\bf Lower right panel:} \textcolor{black}{The black}, green and magenta lines show the values of $\Delta\xi$ in the cases with \textcolor{black}{the} fiducial model $\theta=0$, $\theta=\pi/4$, $\theta=\pi/2$, respectively. In all cases, we have chosen $m_1=10M_{\odot}$, $m_2=1.4M_{\odot}$, $\xi=0$, $d_{\rm L}=10^3$Mpc, $\psi=0$.
}\label{f1}
\end{figure}

Since the gravitational waveform $\tilde{h}(f)$ depends on various angles, including the sky-position angles ($\theta,\phi$), the inclination angle $\iota$ and the polarization angle $\psi$, by numerical  \textcolor{black}{calculations}, we find that the value of $\Delta\xi$ is independent of $\phi$, which is caused by the  equilateral triangle structure of  \textcolor{black}{ET}. However, the dependence of $\Delta\xi$ on the other angle parameters are quite significant. In Fig. \ref{f1}, we consider the case with various angles, which shows that the value of $\Delta\xi$  \textcolor{black}{is minimized} at ($\theta=\iota=\psi=0$). The dependence on $\theta$ and $\iota$ is similar,  \textcolor{black}{and the value of $\Delta\xi$ is maximized}  at $\theta=\pi/2$ and/or $\iota=\pi/2$. On the other hand, the dependence on $\psi$ is quite different. In general, this dependence is very weak. However, in the case  \textcolor{black}{ $\theta=\iota=\pi/2$, i.e.,  when} the orbital plane of  \textcolor{black}{the binary system  is  coincident with the detector plane}, the dependence on $\psi$ \textcolor{black}{becomes} very strong. \textcolor{black}{In particular, when $\psi=(1+2k)\pi/4$ $(k=0,1,2,3)$, it becomes very large, which is caused by the following reason: In the case with $\theta=\iota=\pi/2$ and $\psi=(1+2k)\pi/4$, the pattern functions of ET in Eq. (C6) are $_iF_+(\theta,\phi,\psi)=0$ ($i=1,2,3$), and the cross mode in Eq. (\ref{eq56}) is $h_{\times}(t)=0$. Thus, the leading-order terms of $\tilde{h}(f)$ in Eq. (\ref{hf-total}) become zero, and the parameter constraints are quite loose in this case.}

 \textcolor{black}{Now, let us consider the angle averaged $\Delta\xi$ for the GW bursts. If we consider the restricted PN approximation of the waveform, where all the amplitude corrections of high PN orders are discarded and only PN contributions to the phase are taken into account \cite{restrict}, i.e.
\begin{equation}
\tilde{h}(f)\simeq 2^{-1/2}\frac{M_c^{5/6}}{d_{\rm L}}\sqrt{\frac{5}{48}}\pi^{-2/3}f^{-7/6}Qe^{-i\varphi_{(2,0)}}P_{(2,0)}\Theta(2f_{\rm LSO}-f)\exp[i(2\pi ft_c-\pi/4+2\psi(f/2))],
\end{equation}
analytical calculations show that the mean value $\overline{\Delta\xi}$ obtained by averaging the angles ($\theta,\psi,\iota$) is reduced by a factor $5/2$, compared with
the minimal value of $\Delta\xi_{\min}$ (which is achieved at $\theta=\psi=\iota=0$). However,
in the terms of high PN orders in amplitude, which have not been included in the restricted PN approximation, the dependence of $\tilde{h}(f)$ on the angles ($\theta,\psi,\iota$) are quite complicated through the functions $P_{(m,n)}$ and $\varphi_{(m,n)}$ (see the expressiones of $\tilde{h}^{(k)}(f)$ in Eq. (\ref{hf-total}) and the below one).
So, if taking into account the contributions of these terms, the ratio $\overline{\Delta\xi}/\Delta\xi_{\min}$ deviates from $5/2$ in general.
For given GW detector, the effects of these high PN terms become more significant for the binary system with larger mass of black hole \cite{chris}, which could induce the more significant deviation of the ratio from $5/2$. In order to investigate this kind of derivations, for the binary systems with different masses, we simulate the random samples to compute the values of $\overline{\Delta\xi}$ and compare with the corresponding $\Delta\xi_{\min}$. The results are presented in Table \ref{table1}. As anticipated, we find that if $m_1$ becomes larger, the ratio $\overline{\Delta\xi}/\Delta\xi_{\min}$ becomes more and more deviating from $5/2$. However, this table shows that the value of ratio only slightly deviates from $5/2$. This is in particular the case for $m_1<10M_{\odot}$. So, in general, we can roughly estimate the angle averaged $\Delta\xi$ by the relation $\overline{\Delta\xi}\sim2.5 \Delta\xi_{\min}$.}

\begin{table}
\caption{The numerical ratio $\overline{\Delta\xi}/\Delta\xi_{\min}$ for different cases. The angle averaged value $\overline{\Delta\xi}$ is calculated based on $10^{6}$ random samples for \textcolor{black}{each} case. In each sample, we fix the parameters in the fiducial model
as: $d_{\rm L}=10^3{\rm Mpc}$, $t_c=0$, $\psi_c=0$, $\xi=0.001$ $m_2=1.4M_{\odot}$, $m_1$, and randomly choose the angle parameters $(\theta,\cos\phi,\cos\iota,\psi)$ in the full parameter space.}
\begin{center}
\label{table1}
\begin{tabular}{ c c c c c c }
    \hline
     & ~~$m_1=2M_{\odot}$~~ & ~~$m_1=5M_{\odot}$ ~~& ~~$m_1=10M_{\odot}$ ~~&~~ $m_1=20M_{\odot}$ ~~&~~ $m_1=50M_{\odot}$ ~~ \\
    \hline
    \\
    $\overline{\Delta\xi}/\Delta\xi_{\min}$& $2.498$ & $2.482$ & $2.450$  & $2.325$& $2.151$ \\
   \hline
\end{tabular}
\end{center}
\end{table}

\subsubsection{\textcolor{black}{Dependance on the redshifts} of the sources}

\begin{figure}
\begin{center}
\centerline{\includegraphics[width=15cm]{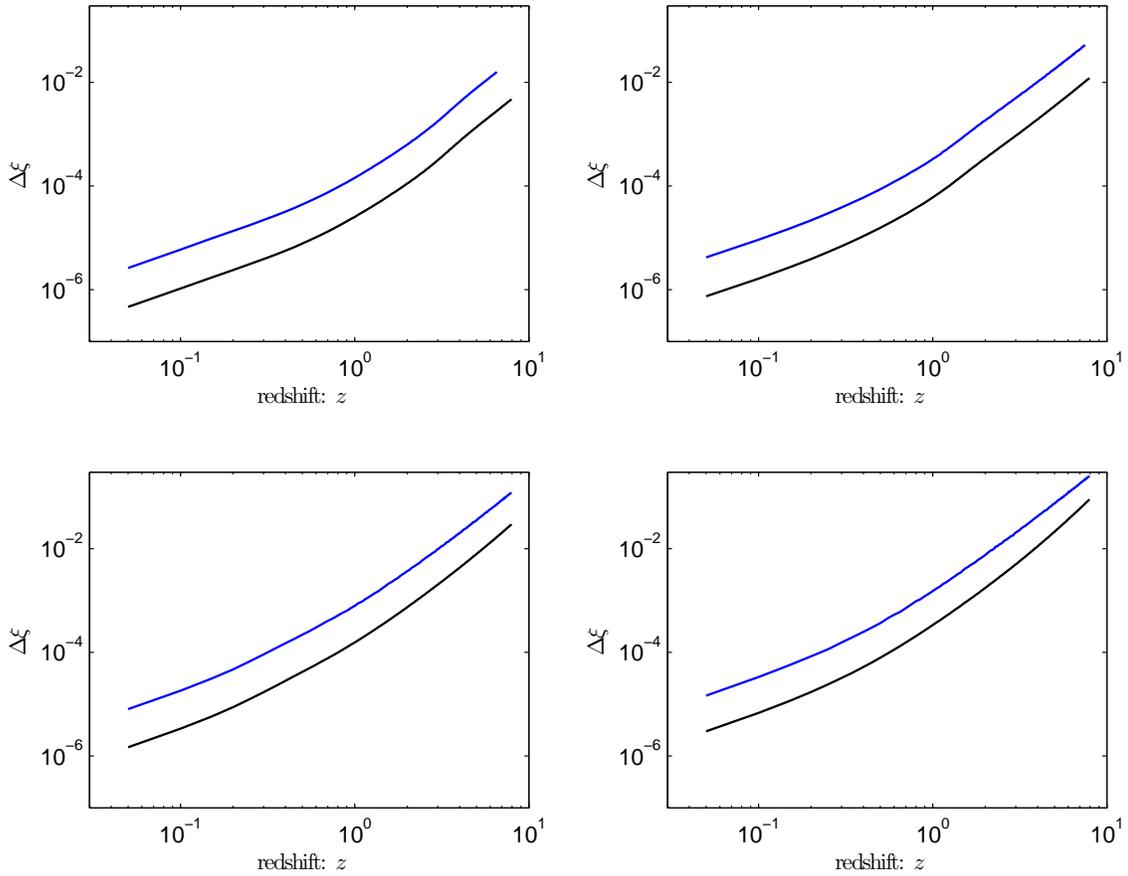}}
\end{center}\caption{ The value of $\Delta\xi$ is determined by the GW bursts at different \textcolor{black}{redshifts} $z$.
In \textcolor{black}{the} upper left panel, we consider the binary system with ($m_{1,{\rm phys}}=2M_{\odot},~m_{2,{\rm phys}}=1.4M_{\odot}$).
In \textcolor{black}{the} upper right panel, we consider the system with ($m_{1,{\rm phys}}=5M_{\odot},~m_{2,{\rm phys}}=1.4M_{\odot}$).
In \textcolor{black}{the}  lower left panel, we consider the system with ($m_{1,{\rm phys}}=10M_{\odot},~m_{2,{\rm phys}}=1.4M_{\odot}$).
In \textcolor{black}{the} lower right panel, we consider the system with ($m_{1,{\rm phys}}=20M_{\odot},~m_{2,{\rm phys}}=1.4M_{\odot}$).
In each panel, the black line shows the results with $(\theta=\psi=\iota=0)$, and the blue line shows the results with $(\theta=\psi=\iota=\pi/2)$.
}\label{f4}
\end{figure}

The redshift $z$ affects the gravitational waveforms by two ways: First, it changes the luminosity distance $d_{\rm L}$. A higher redshift $z$ follows a larger $d_{\rm L}$, which makes the constraint on $\xi$ weaker. Secondly, it \textcolor{black}{changes} the observed masses of the binary system, i.e. $m_{i}=(1+z)m_{i,{\rm phys}}$ ($i=1,2$). A higher $z$ follows a larger $m_{i}$, which also makes the constraint on $\xi$ weaker. Combining these two effects, from Fig. \ref{f4}, we find that the values of $\Delta\xi$ increase about four orders if the redshift of \textcolor{black}{the} GW burst changes from $z=0.05$ to $z=5$. If ET \textcolor{black}{observes} a burst event with $m_{1,{\rm phys}}=2M_{\odot}$ and $m_{2,{\rm phys}}=1.4M_{\odot}$ at redshift $z=0.05$, we expect to obtain a constraint $\Delta\xi\sim 10^{-6}$. If this source is located at $z=1$, the constraint becomes $\Delta\xi\sim 10^{-5}$, and the corresponding constraint on $\omega_{\rm BD}$ is $\omega_{\rm BD}\gtrsim 10^{5}$, which is more stringent than the current upper limit. However, if the event is at $z=5$, the constraint becomes quite loose, i.e. $\Delta\xi\sim 10^{-2}$. So, we expect that the main contribution \textcolor{black}{to} the constraint on $\xi$ comes from the sources in the lowest redshift \textcolor{black}{band}.

\begin{figure}
\begin{center}
\centerline{\includegraphics[width=10cm]{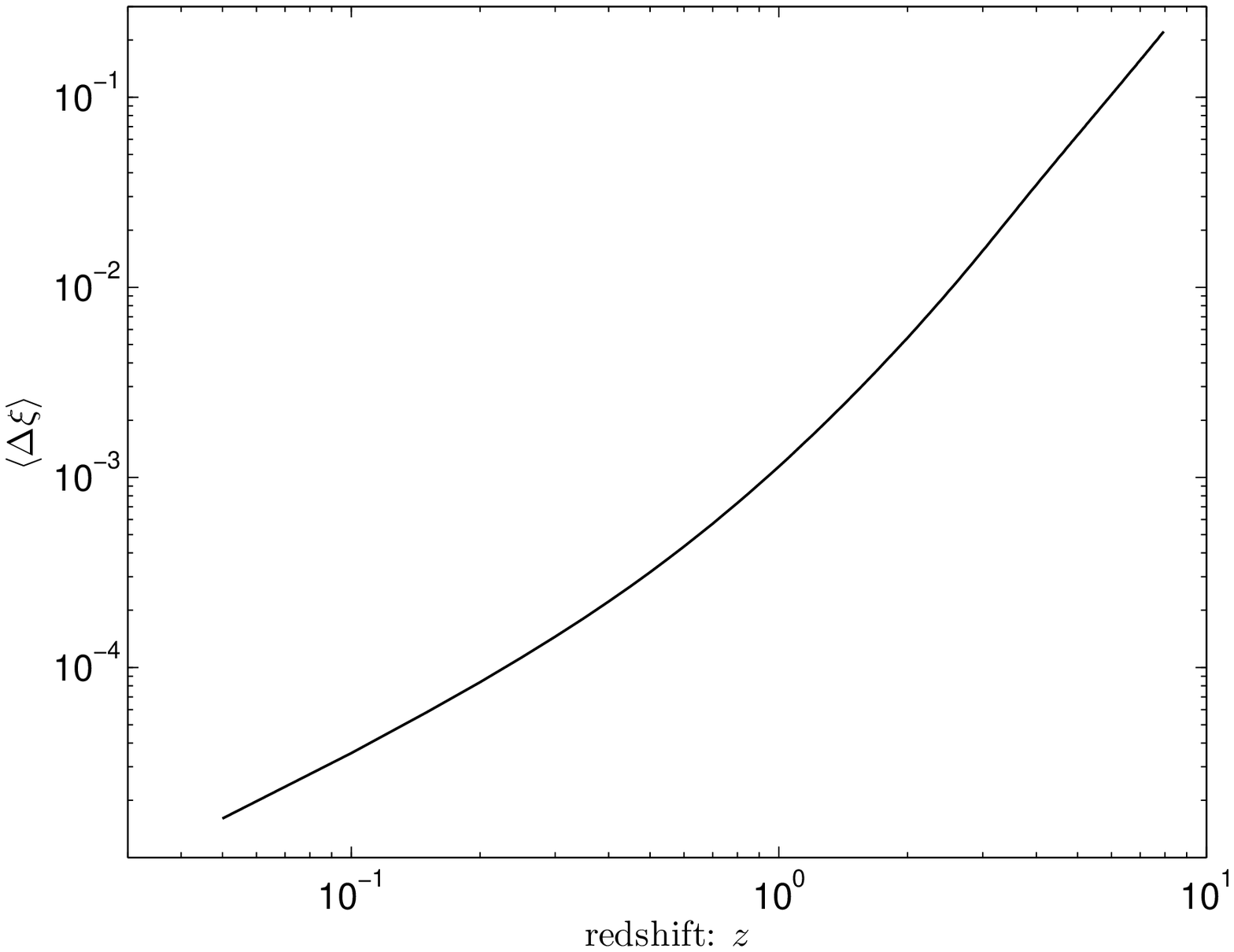}}
\end{center}\caption{ The average value of $\langle{\Delta\xi}\rangle$ for different \textcolor{black}{redshifts} $z$. Note that, $\langle{\Delta\xi}\rangle$
 is computed by averaging the parameters $(m_1,\theta,\psi,\iota)$, where we \textcolor{black}{ take} $m_{1,{\rm phys}}\in[2,100]M_{\odot}$, $\cos\theta\in[-1,1]$, $\psi\in[0,2\pi]$ and $\cos\iota\in[-1,1]$. The other parameters in the fiducial model are given by $m_{2,{\rm phys}}=1.4M_{\odot}$, $\phi=0$, $t_c=0$, $\psi_c=0$ and $\xi=0$.
    }\label{f3}
\end{figure}

For \textcolor{black}{a} given redshift $z$,  we can calculate the averaged value $\langle{\Delta\xi}\rangle$ by taking into account the distribution of black hole \textcolor{black}{masses} $m_{1,{\rm phys}}$, and the angles $(\theta,\phi,\psi,\iota)$. Assuming the uniform distribution of $m_{1,{\rm phys}}$ in the range from $2M_{\odot}$ to $100M_{\odot}$, we plot the results in Fig. \ref{f3}, from which we find that $\langle{\Delta\xi}\rangle=1.6\times 10^{-5}$ for $z=0.05$, and $\langle{\Delta\xi}\rangle=1.1\times 10^{-3}$ for $z=1$. \textcolor{black}{Comparing with the results in Fig. \ref{f4}, we find that for any given redshift $z$, the value of $\langle{\Delta\xi}\rangle$ is much larger than $\Delta\xi$, which is caused by the contribution of higher mass black holes when performing averaging in Fig. \ref{f3}. }

\subsubsection{\textcolor{black}{Dependance} on the total number and distribution of burst events}

The expected rate of \textcolor{black}{coalescences} per year within the horizon of ET is very large for neutron star/neutron star binaries and neutron star/black hole binaries \cite{sayth}. In comparison with the case with a single GW burst event, combining all the events \textcolor{black}{together} can significantly improve the constraint on \textcolor{black}{the} parameter $\xi$. In this subsection, we shall focus on this issue.

For a given cosmological model, the number distribution $f(z)$ of the GW burst events is given by
\bea
f(z)=\frac{4\pi\mathcal{N}r(z)d_{\rm C}^2(z)}{H(z)(1+z)},
\ena
where $d_{\rm C}$ is the comoving distance, which is defined as $d_{\rm C}(z)=\int_0^{z}1/H(z')dz'$ \cite{weinberg}. The function $r(z)$ describes the time evolution of the burst rate, and the constant $\mathcal{N}$ (the number of the sources per comoving volume at redshift $z=0$ over the observation period) is fixed by requiring the total number of the sources $N_{\rm GW}=\int_0^{z_{\max}}f(z)dz$. Actually, the distribution of the events is quite unclear, since \textcolor{black}{ no any  GW burst of this kind
has been detected} until now \cite{s0}. \textcolor{black}{Even for the stable orbiting systems of neutron star/black hole, this has not been} confirmed from observation \cite{s1}. The theoretical estimation shows that the number of neutron star/black hole binary systems should be one or two orders smaller  than  those of neutron star/neutron star \cite{s2}. Since, the
expected total number of inspirals per year within the horizon of ET is \textcolor{black}{about} several $\times 10^{5}$ for neutron star binaries \cite{sayth}, we expect that \textcolor{black}{the} total number of inspiraling neutron star/black hole binaries per year is $\sim 10^{4}$. However, two factors may increase this estimation: First, the discovery of GW bursts, GW150914 and GW151226 may imply that the number of stellar-mass black hole is larger than we expected above \cite{s3}. Second, the neutron star/black hole binaries always emit \textcolor{black}{ stronger GW signals}, so \textcolor{black}{the} detectable distance of this systems is larger than that of binary neutron \textcolor{black}{stars}. Taking into account these factors, for ET, the number of detectable neutron star/black hole \textcolor{black}{binary systems} could be similar \textcolor{black}{to}  that of neutron star/neutron star \cite{s2}.

\begin{figure}
\begin{center}
\centerline{\includegraphics[width=10cm]{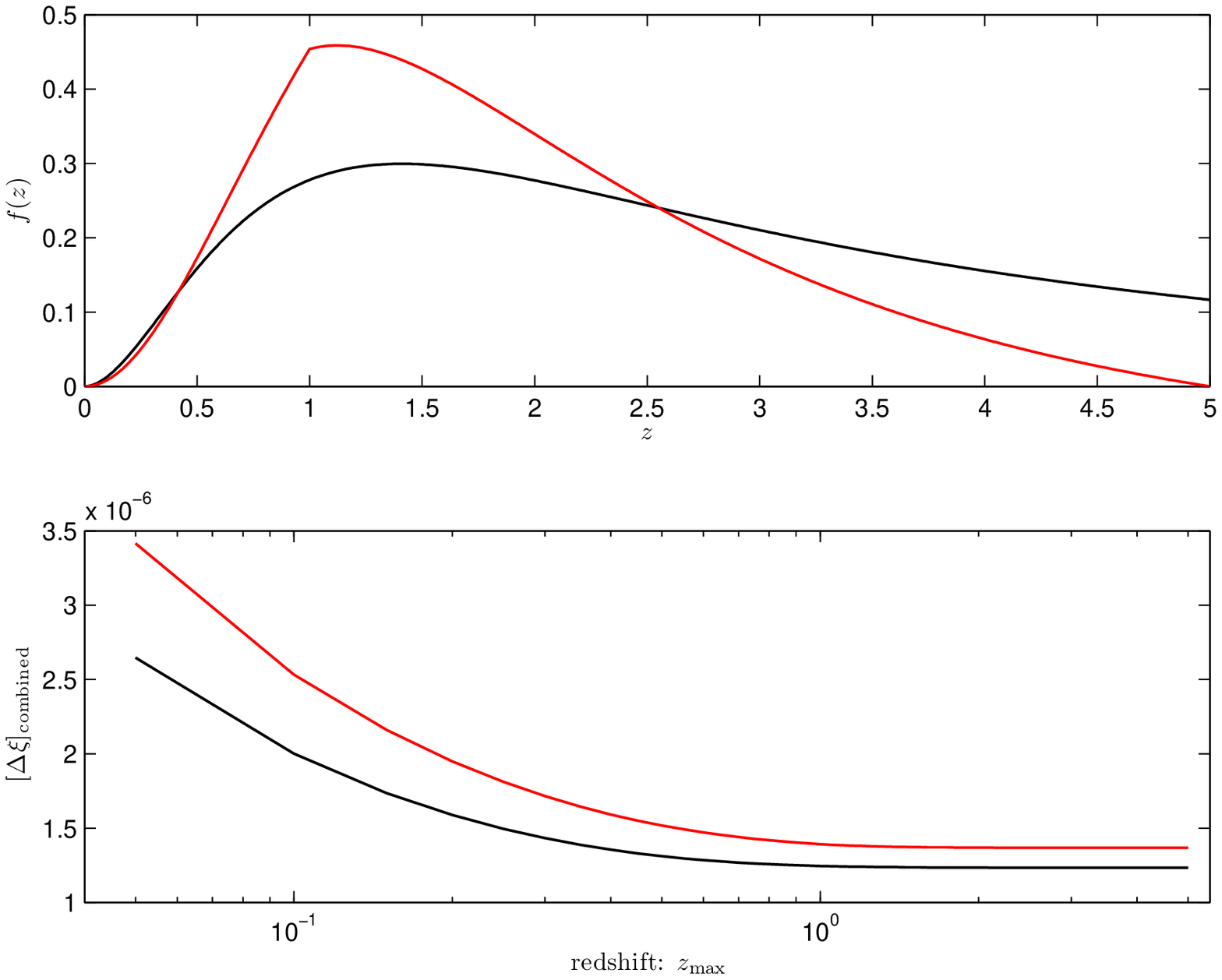}}
\end{center}\caption{{\bf Upper panel:} The normalized distribution of \textcolor{black}{GW} sources in the case with \textcolor{black}{the} uniform distribution (\textcolor{black}{the} black line) and \textcolor{black}{the} nonuniform distribution (\textcolor{black}{the} red line). \\
{\bf Lower panel:} The combined $[\Delta\xi]_{\rm combined}$ by considering all the objects in the redshift range $z\in[0.05,z_{\max}]$.
In this figure, we have assumed $10^4$ observed objects \textcolor{black}{ in} the whole redshift range $z\in[0.05,5]$.
\textcolor{black}{The black and red lines} show the results of \textcolor{black}{the} uniform and nonuniform distributions of the objects, respectively.
}\label{f5}
\end{figure}

In addition to the total number, the time evolution of the source rate is also \textcolor{black}{not clear}. In
this paper we shall consider two different forms for the function
$r(z)$. In the first case we assume that the sources are
distributed uniformly, i.e., with constant comoving number density
throughout the redshift range $0\le z \le 5$ (hereafter we will
refer to this as the uniform distribution). In this case we have
$r(z)=1$. In the other case, we take $r(z)$ to be the following
function: $r(z)=(1+2z)$ for $z\le 1$, $r(z)=(15-3z)/4$ for
$1<z<5$, and $z=0$ for $z\ge 5$. This approximate fit to the rate
evolution is suggested in \cite{nonuniform}. Hereafter, we shall
call this the nonuniform distribution. In the upper panel of Fig. \ref{f5}, we
plot the normalized distribution function $f$ as a function of redshift $z$
in the two cases. Note that in the case with \textcolor{black}{the} nonuniform
distribution, the sources are a little \textcolor{black}{bit} more concentrated at $z=1$.
In what follows we will find out how this affects the
uncertainties on the model parameters.

Considering multiple independent GW burst events, the combined rms error of the parameter $\xi$ can be calculated by
\bea
[\Delta\xi]_{\rm combined}=\left(\sum_{k=1}^{N_{\rm GW}}\frac{1}{[\Delta\xi(k)]^2}\right)^{-1/2}
\ena
where $\Delta\xi(k)$ is the error of $\xi$ derived from the $k$-th source. For the given normalized distribution of the sources, the value of $[\Delta\xi]_{\rm combined}$ depends on the total number $N_{\rm GW}$ \textcolor{black}{through} $\Delta\xi\propto1/\sqrt{N_{\rm GW}}$. In \textcolor{black}{the} lower panel of Fig. \ref{f5}, we plot the combined error of $\xi$ by combining all the objects in the range $z\in[0.05,z_{\max}]$, where we have assumed the total number of events $N_{\rm GW}=10^4$ at $z_{\max}=5$, and the uniform distributions of \textcolor{black}{the} parameters $m_{1,{\rm phys}}\in[2,100]M_{\odot}$, $\cos\theta\in[-1,1]$, $\psi\in[0,2\pi]$ and $\cos\iota\in[-1,1]$. From \textcolor{black}{ this panel}, we find that
\bea
[\Delta\xi]_{\rm combined}=1.23\times10^{-6}\left(\frac{10^4}{N_{\rm GW}}\right)^{1/2}, ~~~{\rm i.e. }~~\omega_{\rm BD}>0.81\times10^{6}\left(\frac{N_{\rm GW}}{10^4}\right)^{1/2},
\ena
for the case with \textcolor{black}{the} uniform distribution. So, even in the conservative case with $N_{\rm GW}=10^4$, the constraint on $\omega_{\rm BD}$ will be 20 times more stringent than the current upper limit derived from the Cassini experiment. If we consider the \textcolor{black}{ case} $N_{\rm GW}=2\times10^5$ observed by ET, the constraint of $\omega_{\rm BD}$ will be improved by two orders compared with the current upper limit. From \textcolor{black}{the} lower panel of Fig. \ref{f5}, we also find that the main contribution comes from the events in \textcolor{black}{the} low frequency range, and the contribution of sources at $z>1$ is ignorable. For the case with \textcolor{black}{the} nonuniform distribution, the constraint becomes
\bea
[\Delta\xi]_{\rm combined}=1.37\times10^{-6}\left(\frac{10^4}{N_{\rm GW}}\right)^{1/2}, ~~~{\rm i.e. }~~\omega_{\rm BD}>0.73\times10^{6}\left(\frac{N_{\rm GW}}{10^4}\right)^{1/2},
\ena
which is slightly weaker than \textcolor{black}{that in }the uniform case, since in the nonuniform case, less events \textcolor{black}{are} distributed \textcolor{black}{in} the low frequency range $z<0.5$.

\section{Conclusions \label{section4}}

The discovery of \textcolor{black}{GW} bursts GW150914 and GW151226 by LIGO opens the new era of \textcolor{black}{the GW} astronomy, where \textcolor{black}{tests of different } theories of gravity in the strong gravitational {fields} is one of the most important \textcolor{black}{issues}. The current observations of \textcolor{black}{the} advanced LIGO have placed interesting constraints on some theories. By ET, a third-generation ground-based GW observatory, the total number and the distance of the observable GW sources, including the inspiralling binary systems, will be greatly improved, which will provide an excellent laboratory to precisely test various gravitational effects, as well as various theories of gravity in the strong gravitational \textcolor{black}{fields}. As \textcolor{black}{an} example, in this paper, we investigate the test ability of ET on \textcolor{black}{BD} gravity by constraining the model parameter $\omega_{\rm BD}$. Up to the lowest PN order, we first calculate the waveforms of gravitational \textcolor{black}{radiations}, including the quadrupole radiation of \textcolor{black}{the} metric field, and the monopole, dipole, quadrupole radiations of \textcolor{black}{the} scalar field, and \textcolor{black}{then} decompose them into the ``plus", ``cross" and ``breathing"
\textcolor{black}{modes}. Employing the stationary phase approximation, we derive the Fourier transforms of these modes, and parameterize the modifications of waveforms in the amplitude, phase and polarization, relative to those in GR. Utilizing the Fisher information matrix, we \textcolor{black}{study} the potential constraints on the parameter $\omega_{\rm BD}$ by ET, and find that \textcolor{black}{an} inspiralling compact binary system composed of a neutron star and a black hole \textcolor{black}{gives the strongest constraints on} BD gravity. The bound on $\omega_{\rm BD}$ depends on the mass of the black hole, the redshift of the system, and the sky-position angle $\theta$, the  inclination angle of binary's orbital $\iota$ and the polarization angle $\psi$. Consistent with the previous \textcolor{black}{results}, we \textcolor{black}{find} that the system with a lower mass can give \textcolor{black}{rise to a} tighter bound on $\omega_{\rm BD}$. If a binary system with \textcolor{black}{a} $2M_{\odot}$ black hole at redshift $z=0.1$ is observed by ET, one expects to obtain a bound $\omega_{\rm BD}\gtrsim O(10^{6})$, which is much more stringent than the current bound derived from the Cassini-Huygens experiment. Combining all the GW burst events can significantly improve the bound, which could arrive at $\omega_{\rm BD}\gtrsim 10^{6}\times(N_{\rm GW}/10^4)^{1/2}$. So, even in the very conservative considerations with
\textcolor{black}{the}  total number of events $N_{\rm GW}=10^4$, the bound is more than one order tighter than the current limit \textcolor{black}{obtained} from Solar System experiments. \textcolor{black}{Hence}, we conclude that the testing ability of ET on
\textcolor{black}{ theories} of gravity is quite promising.


\section*{Acknowledgements}

We appreciate the helpful discussion with Zhoujian Cao. W. Zhao is supported \textcolor{black}{in part} by NSFC \textcolor{black}{Grants Nos.} 11603020, 11633001, 11173021, 11322324, 11653002 \textcolor{black}{and} 11421303, \textcolor{black}{the}  project of Knowledge Innovation Program of Chinese Academy of Science, the Fundamental Research Funds for the Central Universities and the Strategic Priority Research Program of the Chinese Academy of Sciences Grant No. XDB23010200. A. Wang is supported in part by Ciencia Sem Fronteiras, Grant No. A045/2013 CAPES, Brazil, and  NSFC, Grant Nos. 11375153 and 11173021.


\appendix

\section{Decomposing polarization modes of gravitational wave \label{appendixA}}

A generic GW detector measures the local components of \textcolor{black}{the} ``electric" components of the Riemann curvature tensor $R^{0i0j}$, which can \textcolor{black}{be}  formally written as $R^{0i0j}\equiv-(1/2)d^2{\rm h}^{ij}/dt^2$. In general there are six independent components, which can be expressed in terms of polarizations \cite{will-book}. For a wave propagating
in \textcolor{black}{the} $z$-direction, they can be displayed as by the matrix
    \begin{equation}
    {\rm h}_{ij}(t)=\left(
    \begin{matrix}
    {h}_b+{h}_+ & {h}_{\times} & {h}_x \\
    {h}_{\times} & {h}_b-{h}_+ & {h}_y \\
    {h}_x & {h}_{y} & {h}_L
    \end{matrix}
    \right).
    \end{equation}
Three modes ($h_{+}$, $h_{\times}$ and $h_{b}$) are transverse to the direction of propagation, with two ($h_{+}$, $h_{\times}$) representing \textcolor{black}{quadrupole} deformations and one ($h_{b}$) representing a \textcolor{black}{monopole, i.e. the breathing deformation. Three
modes are longitudinal, with one ($h_L$) axially symmetric stretching mode in the propagation
direction, and one quadrupole mode in each of the two orthogonal planes containing the propagation direction ($h_{x}$ and $h_y$).}

Now, let us turn to the \textcolor{black}{GW} ${\rm h}_{ij}(t)$ \textcolor{black}{that} propagates in the
direction $\hat{n}=(1,\theta,\phi)$ in the coordinate system
$X\equiv(x,y,z)$. We first consider the GW in another coordinate system
$X'\equiv(x',y',z')$ with $\hat{n}=\hat{z}'$, where we have \bea
{h}_{+}&=&({\rm h}'_{11}-{\rm h}'_{22})/2,~~{
h}'_{\times}={\rm h}'_{12},  ~~{h}_b=({\rm h}'_{11}+{\rm
h}'_{22})/2,~~{h}_{L}={\rm h}'_{33},  ~~{h}_x={\rm
h}'_{13},~~{h}_{y}={\rm h}'_{23}.\ena
The tensor ${\rm h}'_{ij}$ relates to ${\rm h}_{ij}$ by ${\rm
h}'_{ij}=(R^{T}{\rm h}R)_{ij}$, and the transformation tensor $R$ is given by
    \begin{equation}
    R=\left(
    \begin{matrix}
    \cos\phi & \sin\phi & 0 \\
    -\sin\phi & \cos\phi & 0 \\
    0 & 0 & 1
    \end{matrix}
    \right)\left(
    \begin{matrix}
    1 & 0 & 0 \\
    0 & \cos\theta & \sin\theta \\
    0 & -\sin\theta & \cos\theta
    \end{matrix}
    \right).
    \end{equation}
Thus, we derive the following decompositions \bea {h_{+}}&=&\frac{1}{2}\left\{{\rm
h}_{11}(\cos^2\phi-\cos^2\theta\sin^2\phi)+{\rm
h}_{22}(\sin^2\phi-\cos^2\phi\cos^2\theta)-{\rm
h}_{33}\sin^2\theta \right. \nonumber \\
&&\left.~~~-{\rm h}_{12}[\sin2\phi(1+\cos^2\theta)]+{\rm
h}_{13}\sin\phi\sin2\theta+{\rm
h}_{23}\cos\phi\sin2\theta\right\},
\\
{h}_{\times}&=&\frac{1}{2}\left\{({\rm h}_{11}-{\rm
h}_{22})\cos\theta\sin2\phi+{\rm h}_{12}(2\cos\theta\cos2\phi)-{\rm
h}_{13}(2\sin\theta\cos\phi)+{\rm h}_{23}(2\sin\theta\sin\phi)\right\},
\\
{h}_{b}&=&\frac{1}{2}\left\{{\rm
h}_{11}(\cos^2\phi+\cos^2\theta\sin^2\phi)+{\rm
h}_{22}(\sin^2\phi+\cos^2\phi\cos^2\theta)+{\rm
h}_{33}\sin^2\theta \right. \nonumber \\
&&\left.~~~-{\rm h}_{12}(\sin2\phi\sin^2\theta)-{\rm
h}_{13}\sin\phi\sin2\theta-{\rm
h}_{23}\cos\phi\sin2\theta\right\},
\\
{h}_{L}&=&\frac{1}{2}\left\{{\rm
h}_{11}(2\sin^2\theta\sin^2\phi)+{\rm
h}_{22}(2\cos^2\phi\sin^2\theta)+{\rm
h}_{33}(2\cos^2\theta) \right. \nonumber \\
&&\left.~~~+{\rm h}_{12}(2\sin2\phi\sin^2\theta)+{\rm
h}_{13}(2\sin\phi\sin2\theta)+{\rm
h}_{23}(2\cos\phi\sin2\theta)\right\}, \\
{h}_{x}&=&\frac{1}{2}\left\{({\rm h}_{11}-{\rm
h}_{22})\sin\theta\sin2\phi+{\rm h}_{12}(2\sin\theta\cos2\phi)+{\rm
h}_{13}(2\cos\phi\cos\theta) \right. \nonumber\\
&&~~~\left.-{\rm h}_{23}(2\sin\phi\cos\theta)\right\}, \\
{h}_{y}&=&\frac{1}{2}\left\{{\rm
h}_{11}(\sin2\theta\sin^2\phi)+{\rm
h}_{22}(\cos^2\phi\sin2\theta)-{\rm
h}_{33}(\sin2\theta) \right. \nonumber \\
&&\left.~~~+{\rm h}_{12}(\sin2\phi\sin2\theta)+{\rm
h}_{13}(2\sin\phi\cos2\theta)+{\rm
h}_{23}(2\cos\phi\cos2\theta)\right\}. \ena

\section{Higher post-Newtonian orders of GW waveform in Einstein's General Relativity}

A compact binary system located at the sky-position ($\theta$, $\phi$) with the angle of orbital inclination $\iota$ and polarization angle $\psi$, including the higher PN order terms, the waveforms in the two polarizations \textcolor{black}{are given by},
\bea h_{+,\times}(t)=\frac{2\eta m
x}{d_{\rm L}}\left\{H_{+,\times}^{(0)}+x^{1/2}H_{+,\times}^{(1/2)}+xH_{+,\times}^{(1)}+x^{3/2}H_{+,\times}^{(3/2)}+x^{2}H_{+,\times}^{(2)}+x^{5/2}H_{+,\times}^{(5/2)}+O({1}/{c^6})\right\},\ena
where $m=m_1+m_2$ is the total mass, $\eta=m_1m_2/m^2$ is the symmetric mass ratio. The PN expansion parameter is defined as $x\equiv v^2$. The coefficients $H^{(i/2)}_{+,\times}$ ($i=0,1,\cdot\cdot\cdot,5$), are linear combinations of various harmonics with prefactors that depend on $\iota$ and $\eta$. The lowest order ones are,
\bea
H^{(0)}_{+}&=&-(1+\cos^2\iota)\cos2\Phi(t)-(1/96)\sin^2\iota(17+\cos^2\iota),\\
H^{(0)}_{\times}&=&-2\cos\iota\sin2\Phi(t),\\
\Phi(t)&=&\phi(t)-2m\omega_s\ln(\omega_s/\omega_0),\ena
\textcolor{black}{where $\omega_0$ is a constant frequency that can be conveniently chosen as the entry frequency of an interferometric detector \cite{gw-book}.} The other
terms can be found in the previous work \cite{amplitude}.

In general, we can write them as \bea
H_{+,\times}^{(s)}=\sum_{n}\left\{C_{+,\times}^{(n,s)}\cos[n\Phi(t)]+D_{+,\times}^{(n,s)}\sin[n\Phi(t)]\right\}\ena
Thus, we have \bea
h(t)&\equiv&F_+h_+(t)+F_{\times}h_{\times}(t)=\frac{2\mu
x}{d_{\rm L}}\sum_{n,s}x^s\left\{C^{(n,s)}\cos[n\Phi(t)]+D^{(n,s)}\sin[n\Phi(t)]\right\}
\\
&=&\frac{2\mu
x}{d_{\rm L}}\sum_{n,s}\left\{x^sP_{(n,s)}e^{i[n\psi+\varphi_{(n,s)}]}\right\}\ena
where \bea
C^{(n,s)}&=&F_+C_+^{(n,s)}+F_{\times}C_{\times}^{(n,s)},~~~D^{(n,s)}=F_+D_+^{(n,s)}+F_{\times}D_{\times}^{(n,s)},\\
P_{(n,s)}&=&{\rm
sign}[F_+C_{+}^{(n,s)}+F_{\times}C_{\times}^{(n,s)}]\left\{[F_+C_{+}^{(n,s)}+F_{\times}C_{\times}^{(n,s)}]^2+[F_+D_{+}^{(n,s)}+F_{\times}D_{\times}^{(n,s)}]^2\right\}^{1/2},\\
\varphi_{(n,s)}&=&\tan^{-1}\left\{-\frac{F_+D_{+}^{(n,s)}+F_{\times}D_{\times}^{(n,s)}}{F_+C_{+}^{(n,s)}+F_{\times}C_{\times}^{(n,s)}}\right\}.\ena

The Fourier components of $h(t)$ are given by \bea \tilde
h(f)=\sum_{k=1}^{7}\tilde{h}^{(k)}(f),\ena
where the harmonics are explicitly presented in \cite{chris}, {\emph{e.g.}}, the term $\tilde{h}^{(1)}(f)$ is given by
\bea
\tilde{h}^{(1)}(f)&=&\frac{M_c^{5/6}}{d_{\rm L}}\sqrt{\frac{5}{48}}\pi^{-2/3}(2f)^{-7/6}\left\{e^{-i\varphi_{(1,1/2)}}P_{(1,1/2)}(2\pi
mf)^{1/3}\right. \nonumber \\
&&+\left[e^{-i\varphi_{(1,3/2)}}P_{(1,3/2)}+e^{-i\varphi_{(1,1/2)}}P_{(1,1/2)}S_1\right](2\pi
mf) \nonumber \\
&&+\left[e^{-i\varphi_{(1,2)}}P_{(1,2)}+e^{-i\varphi_{(1,1/2)}}P_{(1,1/2)}S_{3/2}\right](2\pi
mf)^{4/3} \nonumber \\
&&\left.+\left[e^{-i\varphi_{(1,5/2)}}P_{(1,5/2)}+e^{-i\varphi_{(1,3/2)}}P_{(1,3/2)}S_{1}+e^{-i\varphi_{(1,1/2)}}P_{(1,1/2)}S_{2}\right](2\pi
mf)^{5/3}\right\} \nonumber\\
&&\times\Theta(f_{\rm LSO}-f)\exp[i(2\pi ft_c-\pi/4+\psi(f))] .\nonumber\ena
in which \bea
S_1&=&\frac{1}{2}\left(\frac{743}{336}+\frac{11}{4}\eta\right),~S_{3/2}=-2\pi,
~
S_2=\frac{7266251}{8128512}+\frac{18913}{16128}\eta+\frac{1379}{1152}\eta^2.
\nonumber \ena
The phase function is
\bea \psi(f)=-\psi_c+\frac{3}{256(2\pi
M_cf)^{5/3}}\sum_{i=0}^{7}\psi_i(2\pi mf)^{i/3},\ena
where
\bea
\psi_0&=&1,~~
\psi_1=0, ~~
\psi_2=\frac{20}{9}\left[\frac{743}{336}+\frac{11}{4}\eta\right], ~~
\psi_3=-16\pi, ~~
\psi_4=10\left[\frac{3058673}{1016064}+\frac{5429}{1008}\eta+\frac{617}{114}\eta^2\right], \nonumber\\
\psi_5&=&\pi\left[\frac{38645}{756}+\frac{38645}{252}\ln\left({f}/{f_{\rm LSO}}\right)-\frac{65}{9}\eta\left(1+3\ln\left({f}/{f_{\rm LSO}}\right)\right)\right], \nonumber\\
\psi_6&=&\left(\frac{11583231236531}{4694215680}-\frac{640\pi^2}{3}-\frac{6846\gamma}{21}\right)+\eta\left(-\frac{15335597827}{3048192}+\frac{2255\pi^2}{12}-\frac{1760\theta}{3}+\frac{12320\lambda}{9}\right) \nonumber \\
&+&\frac{76055}{1728}\eta^2-\frac{127825}{1296}\eta^3-\frac{6848}{21}\ln[4(2\pi m f)^{1/3}]
, \nonumber\\
\psi_7&=&\pi\left(\frac{77096675}{254016}+\frac{378515}{1512}\eta-\frac{74045}{756}\eta^2\right), \nonumber
\ena
in which $\gamma=0.5772$ is the Euler-Mascheroni constant, $\lambda=-0.6451$ and $\theta=-1.28$.

\section{Pattern functions of Einstein Telescope}

A gravitational wave with a given propagation direction $\hat{\bf n}$ can be written as
\bea
h_{ij}(t,{\bf x})=\sum_{A}e^{A}_{ij}(\hat{\bf n})\int_{-\infty}^{\infty} df\tilde{h}_{A}(f) e^{-2\pi i f(t-\hat{\bf n}\cdot{\bf x})}
\ena
where $A=+,\times,b,L,x,y$, $e_{ij}^{A}$ are the polarization tensors. 
We take ${\bf x}=0$ as the location of the detector. For a detector which is sensitive only to GWs with a reduced wavelength much larger than its size, such as resonant masses and ground-based interferometers, we have $2\pi f\hat{\bf n}\cdot{\bf x}\ll 1$ over the whole detector, and we can neglect the spatial dependence of $h_{ij}(t,{\bf x})$. So, to study the  interaction of GWs with such detectors we can  simply  write
\bea
h_{ij}(t)=\sum_{A}e_{ij}^{A}(\hat{\bf n})\int_{-\infty}^{\infty} df\tilde{h}_{A}(f) e^{-2\pi i ft}=\sum_{A}e_{ij}^{A}(\hat{\bf n})h_{A}(t).
\ena

\begin{figure}
\begin{center}
\centerline{\includegraphics[width=10cm]{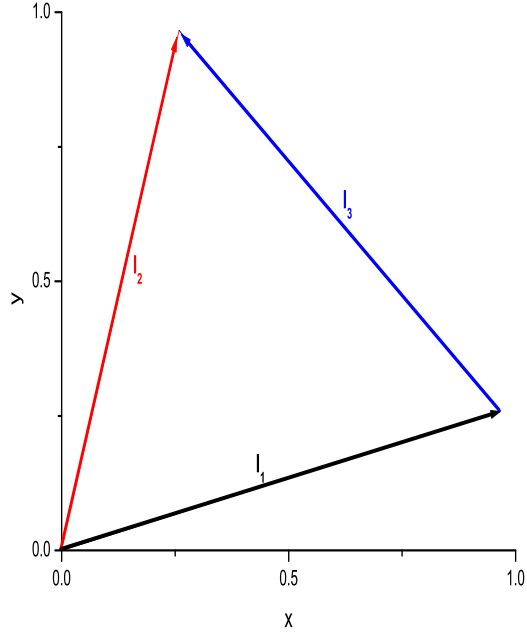}}
\end{center}\caption{Unit vectors defining the detector tensors for a triangular Einstein Telescope.}\label{ET}
\end{figure}

In general, the input of the GW detector has the form
\bea
h(t)=D^{ij}h_{ij}(t)=\sum_{A} D^{ij} e^{A}_{ij}(\hat{\bf n})h_A(t)=\sum_{A}F_{A}(\hat{\bf n})h_A(t),
\ena
where $D^{ij}$ is a constant tensor which depends on the detector geometry, and is known as the detector tensor. $F_{A}(\hat{\bf n})\equiv D^{ij}e^{A}_{ij}(\hat{\bf n})$ is the detector pattern \textcolor{black}{functions}.

Now, let us focus on the \textcolor{black}{ET}. One possible set-up for ET would be a triangular tube with 10 km edges containing three interferometers with 60 degree opening angles. Consider three interferometers with 60 degree opening angles, arranged in an equilateral triangle. Let $\hat{l}_{\rm A}~(A=1,2,3)$ be unit vectors tangent to the edges of the triangles as \textcolor{black}{shown}  in Fig. \ref{ET}. These can be expressed in terms of \textcolor{black}{the unit vectors} $(\hat{x},\hat{y},\hat{z})$ defining a Cartesian coordinate system, where $(\hat{x},\hat{y})$ are in the detector plane:
\bea
\hat{l}_{\rm A}=\cos(\alpha_{\rm A})\hat{x}+\sin(\alpha_{\rm A})\hat{y},
\ena
with $\alpha_{\rm A}=\pi/12+(A-1)\pi/3$. The three interferometers inside the triangular tube have detector tensors
\bea
_1D^{ij}=\frac{1}{2}(\hat{l}_1^{i}\hat{l}_1^{j}-\hat{l}_2^{i}\hat{l}_2^{j}),~~
_2D^{ij}=\frac{1}{2}(\hat{l}_2^{i}\hat{l}_2^{j}-\hat{l}_3^{i}\hat{l}_3^{j}),~~
_3D^{ij}=\frac{1}{2}(\hat{l}_1^{i}\hat{l}_1^{j}-\hat{l}_3^{i}\hat{l}_3^{j}),
\ena
where $i=1,2,3$ are spatial indices.

Assume the GW source in the direction $\hat{\bf n}=(1,\theta,\phi)$ with the polarization angle $\psi$ in the coordinate system $(\hat{x},\hat{y},\hat{z})$. Utilizing the transformation between this system and the coordinate system $(\hat{x}',\hat{y}',\hat{z}')$, we \textcolor{black}{find} that
\bea
_1F_{+}(\theta,\phi,\psi)&=&\frac{\sqrt{3}}{2}\left[\frac{1}{2}(1+\cos^2\theta)\cos2\phi\cos2\psi-\cos\theta\sin2\phi\sin2\psi\right], \\
_1F_{\times}(\theta,\phi,\psi)&=&\frac{\sqrt{3}}{2}\left[\frac{1}{2}(1+\cos^2\theta)\cos2\phi\sin2\psi+\cos\theta\sin2\phi\cos2\psi\right], \\
_1F_{b}(\theta,\phi,\psi)&=&\frac{\sqrt{3}}{2}\left[-\frac{1}{2}\sin^2\theta\cos2\phi\right], \\
_1F_{L}(\theta,\phi,\psi)&=&\frac{\sqrt{3}}{2}\left[\frac{1}{2}\sin^2\theta\cos2\phi\right], \\
_1F_{x}(\theta,\phi,\psi)&=&\frac{\sqrt{3}}{2}\left[\frac{1}{2}\sin2\theta\cos2\phi\cos\psi-\sin\theta\sin2\phi\sin\psi\right], \\
_1F_{y}(\theta,\phi,\psi)&=&\frac{\sqrt{3}}{2}\left[\frac{1}{2}\sin2\theta\cos2\phi\sin\psi+\sin\theta\sin2\phi\cos\psi\right], \\
_2F_{A}(\theta,\phi,\psi)&=&_1F_{A}(\theta,\phi+2\pi/3,\psi),\\
_3F_{A}(\theta,\phi,\psi)&=&_1F_{A}(\theta,\phi+4\pi/3,\psi).
\ena








\baselineskip=12truept

\begin{thebibliography}{99}


\bibitem{will-book}
C. M. Will, {\em Theory and experiment in gravitational physics}, (Cambridge University Press, Cambridge, 1993).



\bibitem{will-review}
C. M. Will, Living Reviews in Relativity {\bf 17}, 4 (2014).



\bibitem{Clifton2012p1}
T. Clifton, P.~G. Ferreira, A. Padilla, and C. Skordis, Phys. Rept. {\bf 513}, 1  (2012).

\bibitem{Stairs2003p5}
I.~H. Stairs, Living Reviews in Relativity {\bf 6},  5  (2003).

\bibitem{Wex2014}
N. Wex,  {\em Testing Relativistic Gravity with Radio Pulsars}, to appear in the Brumberg Festschrift edited by S. M. Kopeikein (de Gruyter, Berlin, to be published) [arXiv:1402.5594].



\bibitem{gw150914}
The LIGO Scientific Collaboration and the Virgo Collaboration, Phys. Rev. Lett. {\bf 116}, 061102 (2016).


\bibitem{yunes2016}
N. Yunes, K. Yagi and F. Pretorius, Phys. Rev. D. {\bf 94}, 084002 (2016).

\bibitem{arzaon2016}
M. Arzano and G. Calcagni, Phys. Rev. D {\bf 93}, 124065 (2016).

\bibitem{maselli}
A. Maselli, S. Marassi, V. Ferrari, K. Kokkotas and R. Schneider, Phys. Rev. Lett. {\bf 117}, 091102 (2016).


\bibitem{karl}
K. Popper, \emph{The Logic of Scientific Discovery}, 2nd ed. (Routledge Press, London and New York, 2002).



\bibitem{ppe}
N. Yunes and F. Pretorius, Phys. Rev. D {\bf 80}, 122003 (2009); N. Loutrel, N. Yunes and F. Pretorius, Phys. Rev. D {\bf 90}, 104010 (2014).


\bibitem{bd-book1}
Y. Fujii and K. Maeda, {\emph{The scalar-tensor theory of gravitation}}, (Cambridge University Press, Cambridge, 2003).

\bibitem{bd-book2}
V. Faraoni, {\emph{Cosmology in scalar-tensor gravity}}, (Kluwer Academic Publishers, London, 2004).


\bibitem{will1994}
C. M. Will,  Phys. Rev. D {\bf 50}, 6058 (1994).


\bibitem{will1989}
C. M. Will and H. W. Zaglauer, Astrophys. J. {\bf 346}, 366 (1989).

\bibitem{lang2013}
R. N. Lang, Phys. Rev. D {\bf 89}, 084014 (2014).

\bibitem{lang2014}
R. N. Lang, Phys. Rev. D {\bf 91}, 084027 (2015).


\bibitem{bd-higher}
S. Mirshekari and C. M. Will, Phys. Rev. D {\bf 87}, 084070 (2013).



\bibitem{Sennett:2016klh}
N. Sennett, S, Marsat and A. Buonanno, Phys. Rev. D {\bf 94}, 084003 (2016).




\bibitem{damour}
T. Damour and G. Esposito-Farese, Class. Quantum Grav. {\bf 9},  2093  (1992).


\bibitem{will-massive}
J. {Alsing}, E. {Berti}, C.~M. {Will}, and H. {Zaglauer}, Phys. Rev. D {\bf 85}, 064041  (2012).

\bibitem{barx2014}
P. Brax, A. C. Davis, and J. Sakstein, Class. Quantum Grav. {\bf 31},  225001 (2014).

\bibitem{cao}
Z. Cao, P. Galaviz and L. Li, Phys. Rev. D {\bf 87}, 104029 (2013).

\bibitem{zhang}
X. Zhang, W. Zhao, H. Huang and Y. Cai, Phys. Rev. D {\bf 93}, 124003 (2016);
\textcolor{black}{X. Zhang, T. Liu and W. Zhao, Phys. Rev. D {\bf 95}, 104027 (2017).}

\bibitem{bound}
D. Bertotti, L. Iess and P. Tortota, Nature {\bf 425}, 374 (2003).



\bibitem{will-lisaa}
P. D. Scharre and C. M. Will, Phys. Rev. D {\bf 65}, 042002 (2002).




\bibitem{will-lisa}
C. M. Will and N. Yunes, Class. Quantum Grav. {\bf 21}, 4367 (2004).


\bibitem{lisa2}
E. Berti, A. Buonanno and C. M. Will, Phys. Rev. D {\bf 71}, 084025 (2005).


\bibitem{lisa3}
K. Yagi and T. Tanaka, Phys. Rev. D {\bf 81}, 064008 (2010).

\bibitem{bbo-decigo}
K. Yagi and T. Tanaka, Prog. Theor. Phys. {\bf 123}, 1069 (2010).




\bibitem{et} ``The Einstein Telescope Project", \texttt{https://www.et-gw.eu/}.


\bibitem{et2}
M. Abernathy et al., {\emph{Einstein Gravitational Wave Telescope: Conceptual Design Study,}}
Document No. ET-0106A-10.

\bibitem{et3}
K. G. Arun and A. Pai,  Int. J. Mod. Phys. D {\bf 22}, 1341012 (2013).



\bibitem{will-book2}
E. Poisson and C. M. Will, {\emph{Gravity}}, (Cambridge University Press, Cambridge, 2014).

\bibitem{eardley}
D.~M. {Eardley}, Astrophys. J. Lett. {\bf 196},  L59  (1975).




\bibitem{hawking1972}
S. W. Hawking, Commun. Math. Phys. {\bf 25}, 167 (1972).

\bibitem{weinberg}
S. Weinberg, {\emph{Cosmology}}, (Oxford University Press, Oxford, 2008).


\bibitem{planck}
Planck Collaboration, Astronomy and Astrophysics, {\bf 571}, A31 (2014).

\bibitem{MTW}
C. W. Misner, K. S. Thorne and J. A. Wheeler, {\em Gravitation}, (W. H. Freeman and Company, Newyork, 1973).



\bibitem{gw-book}
M. Maggiore, {\em Gravitational Waves. Vol. 1: Theory and Experiments}, (Oxford University Press, Oxford, England, 2007).


\bibitem{yunes2012}
K. Chatziioannou, N. Yunes and N. Cornish, Phys. Rev. D {\bf 86}, 022004 (2012).

\bibitem{triassintes}
C. Cutler, Phys. Rev. D {\bf 57}, 7089 (1998);
M. Trias and A. M. Sintes, Phys. Rev. D {\bf 77}, 024030 (2008).


\bibitem{pn-review}
L. Blanchet, Living Reviews in Relativity {\bf 5}, 3 (2002).

\bibitem{amplitude}
L. Blanchet, B. R. Iyer, C. M. Will and A. G. Wiseman, Class. Quantum Grav. {\bf 13}, 575 (1996);
K. G. Arun, L. Blanchet, B. R. Iyer, M. S. S. Qusailah, Class. Quantum Grav. {\bf 21}, 3771 (2004); Erratum ibid. {\bf 22}, 3115 (2005).

\bibitem{phase}
T. Damour, P. Jaranowski, and G. Sch\"afer, Phys. Lett. B {\bf
513}, 147 (2001); Y. Itoh, T. Futamase, and H. Asada, Phys. Rev. D
{\bf 63}, 064038 (2001); L. Blanchet, G. Faye, B. R. Iyer, and B.
Joguet, Phys. Rev. D {\bf 65}, 061501(R) (2002); Erratum ibid. D
{\bf 71}, 129902 (2005); Y. Itoh and T. Futamase, Phys. Rev. D
{\bf 68}, 121501(R) (2003); Y. Itoh, Phys. Rev. D {\bf 69}, 064018
(2004); L. Blanchet, T. Damour, and G. Esposito-Far\`ese, Phys.
Rev. D {\bf 69}, 124007 (2004); L. Blanchet, T. Damour, G.
Esposito-Far\`ese, and B. R. Iyer, Phys. Rev. Lett. {\bf 93},
091101 (2004); Y. Itoh, Class. Quantum Grav. {\bf 21}, S529
(2004); L. Blanchet and B. R. Iyer, Phys. Rev. D {\bf 71}, 024004
(2005).





\bibitem{chris}
C. Van Den Broeck and A. S. Sengupta, Class. Quantum Grav. {\bf 24}, 155 (2007).


\bibitem{zhao2011}
W. Zhao, C. Van Den Broeck, D. Baskaran and T. G. F. Li, Phys. Rev. D {\bf 83}, 023005 (2011).

\bibitem{sayth}
B. S. Sathyaprakash, B. Schutz and C. Van Den Broeck, Class. Quantum Grav. {\bf 27}, 215006 (2010).


\bibitem{gair2}
S. R. Taylor and J. R. Gair, Phys. Rev. D {\bf 86}, 023502 (2012).

\bibitem{cai}
R. Cai and T. Yang, Phys. Rev. D {\bf 95}, 044024 (2017).

\bibitem{zhu}
A. Piorkowska, M. Biesiada and Z. Zhu, JCAP {\bf 10}, 022 (2013);
M. Biesiada, X. Ding, A. Piorkowska and Z. Zhu, JCAP {\bf 10}, 080 (2014);
X. Ding, M. Biesiada and Z. Zhu, JCAP {\bf 12}, 006 (2015).

\bibitem{mock}
T. Regimbau, et al., Phys. Rev. D {\bf 86}, 122001 (2012);
T. Regimbau, D. Meacher and M Coughlin, Phys. Rev. D {\bf 89}, 084046 (2014);
D. Meacher, K. Cannon, C. Hanna, T. Regimbau and B. S. Sathyaprakash, Phys. Rev. D {\bf 93}, 024018 (2016).

\bibitem{gair}
J. R. Gair, I. Mandel, M. C. Miller and M. Volonteri, Gen. Rel. Grav. {\bf 43}, 485 (2011);
E. A. Huerta and J. R. Gair, Phys. Rev. D {\bf 83}, 044021 (2011).

\bibitem{et-science}
B. Sathyaprakash, et al., arXiv:1108.1423;
B. Sathyaprakash, et al., Class. Quantum Grav. {\bf 29}, 124013 (2012);
C. Van Den Broeck, Journal of Physics: Conference Series, {\bf 484}, 012008 (2014).


\bibitem{d1}
C. K. Mishra, K. G. Arun, B. R. Iyer and B. S. Sathyaprakash,  Phys.  Rev. D {\bf 82},  064010  (2010).

\bibitem{d2}
W. D. Pozzo, T. G. F. Li and C. Messenger, Phys. Rev. D {\bf 95}, 043502 (2017).


\bibitem{d3}
C. Messenger and J. Read, Phys. Rev. Lett. {\bf 108}, 091101 (2012).



\bibitem{vilta}
\textcolor{black}{S. Vitale and M. Evans, Phys. Rev. D {\bf 95}, 064052 (2017).}


\bibitem{zhu2017}
X. Fan, K. Liao, M. Biesiada, A. Piorkowska-Kurpas and Z. Zhu, Phys. Rev. Lett. {\bf 118}, 091102 (2017).
K. Liao, X. Fan, X. Ding, M. Biesiada and Z. Zhu, arXiv:1703.04151.

\bibitem{freise}
A. Freise, S. Hild, K. Somiya, K. A. Strain, A. Vicere, M. Barsuglia and S. Chelkowski, Gen. Rel. Grav. {\bf 43}, 537 (2011).



\bibitem{gwreview}
B. S. Sathyaprakash and B. F. Schutz, Living Reviews in Relativity {\bf 12}, 2
(2009).

\bibitem{gwfisher}
L. S. Finn, Phys. Rev. D {\bf 46}, 5236 (1992); L. S. Finn and D.
F. Chernoff, Phys. Rev. D {\bf 47}, 2198 (1993).











\bibitem{gamma-ray}
E. Nakar, Phys. Rept. {\bf 442}, 166 (2007).

\bibitem{schutz}
B. Schutz, Nature (London), {\bf 323}, 310 (1986).



\bibitem{restrict}
C. Cutler, et al., Phys. Rev. Lett. {\bf 70}, 2984 (1993).


\bibitem{s0}
The LIGO Scientific Collaboration and the Virgo Collaboration, Astrophys. J. {\bf 832}, L21 (2016).

\bibitem{s1}
J. G. Martinez, et al., Astrophys. J. {\bf 812}, 143 (2015).

\bibitem{s2}
J. Abadie, et al., Class. Quantum Grav. {\bf 27}, 173001 (2010).



\bibitem{s3}
The LIGO Scientific Collaboration and the Virgo Collaboration, Astrophys. J. Suppl. {\bf 227}, 14 (2016).

\bibitem{nonuniform}
R. Schneider, V. Ferrari, S. Matarrese and S. F. P. Zwart, \MNRAS ~{\bf 324}, 797 (2001).



\end{thebibliography}
\end{document}